\documentclass[manuscript]{aastex}

\usepackage{graphicx}        % For eps figures, newer & more powerful
\usepackage[T1]{fontenc}        % Needed for typing Polish ogonek symbols (\k{})

\slugcomment{Submitted to Astrophysical Journal.}
\shorttitle{Ca and Ar X-ray Flare Spectra}
\shortauthors{Phillips et al.}
\begin{document}


%\title{Highly Ionized Calcium X-ray Spectra from Solar Flares}
\title{HIGHLY IONIZED CALCIUM AND ARGON X-RAY SPECTRA FROM A LARGE SOLAR FLARE}

%% Use \author, \affil, plus the \and command to format author and affiliation
%% information.  If done correctly the peer review system will be able to
%% automatically put the author and affiliation information from the manuscript
%% and save the corresponding author the trouble of entering it by hand.
%%
%% The \affil should be used to document primary affiliations and the
%% \altaffil should be used for secondary affiliations, titles, or email.

%% Authors with the same affiliation can be grouped in a single
\author{K. J. H. Phillips\altaffilmark{1}, J. Sylwester\altaffilmark{2}, B. Sylwester\altaffilmark{2}, M. Kowali\'{n}ski\altaffilmark{2}, M. Siarkowski\altaffilmark{2}, \\
W. Trzebi\'{n}ski\altaffilmark{2}, S. P{\l}ocieniak\altaffilmark{2}, Z Kordylewski\altaffilmark{2}}
\affil{$^1$ Earth Sciences Department, Natural History Museum, London SW7 5BD, UK}
\email{kennethjhphillips@yahoo.com}
\affil{$^2$ Space Research Centre, Polish Academy of Sciences, Kopernika 11, 51-622 Wroc{\l}aw, and Bartycka 18A, 00-716 Warszawa, Poland}
\email{bs@cbk.pan.wroc.pl,js@cbk.pan.wroc.pl}

% FOOTNOTE: \latex\ \footnote{\url{http://www.latex-project.org/}}

\begin{abstract}
X-ray lines of helium-like calcium (\ion{Ca}{19}) between 3.17~\AA\ and 3.21~\AA\ and associated \ion{Ca}{18} dielectronic satellites have previously been observed in solar flare spectra, and their excitation mechanisms are well established. Dielectronic satellites of lower ionization stages (\ion{Ca}{17}~--~\ion{Ca}{15}) are not as well characterized. Several spectra during a large solar flare in 2001 by the DIOGENESS X-ray spectrometer on the {\em CORONAS-F}\/ spacecraft show the \ion{Ca}{17} and \ion{Ca}{16} satellites as well as lines of ionized argon (\ion{Ar}{17}, \ion{Ar}{16}) including dielectronic satellites. The DIOGENESS spectra are compared with spectra from a synthesis code developed here based on an isothermal assumption with various atomic sources including dielectronic satellite data from the Cowan Hartree--Fock code. Best-fit comparisons are made by varying the temperature as the code's input (Ar/Ca abundance ratio fixed at 0.33); close agreement is achieved although with adjustments to some ion fractions. The derived temperature is close to that derived from the two {\em GOES}\/ X-ray channels, $T_{\rm GOES}$. Some lines are identified for the first time. Similar spectra from the {\em P78-1}\/ spacecraft and the Alcator C-Mod tokamak have also been analyzed and similar agreement obtained. The importance of blends of calcium and argon lines is emphasized, affecting line ratios used for temperature diagnostics. This analysis will be applied to the {\em Solar Maximum Mission}\/ Bent Crystal Spectrometer archive and to X-ray spectra expected from the ChemiX instrument on the Sun-orbiting {\em Interhelioprobe}\/ spacecraft, while the relevance to X-ray spectra from non-solar sources is indicated.
\end{abstract}




% \ion{Ca}{19}
% ~--~

% THIS VERSION: {\today}

% Section 1
\section{INTRODUCTION} \label{sec:intro}

Highly ionized calcium X-ray spectra have been observed by several instruments on solar spacecraft, particularly those dedicated to the study of high-temperature flare-produced plasmas. Lines due to helium-like Ca (\ion{Ca}{19}) in the wavelength range 3.17~--~3.21~\AA\ occur in flares with {\em GOES}\/ classifications C1 and higher, and with temperatures of up to $\sim 20$~MK. \ion{Ca}{19} spectra were recorded by SOLFLEX ({\em Solar Flare X-rays}) on {\em P78-1}\/ (operating 1979~--~1981), the Bent Crystal Spectrometer on {\em Solar Maximum Mission}\/ (1980 and 1984~--~1989), and the Bragg Crystal Spectrometer on {\em Yohkoh}\/ (1991~--~2001), time periods coinciding with the high-activity maxima of Solar Cycles~21 and 22, when unlike much of the most recent cycle (24) {\em GOES}\/ class M and X flares were commonplace. The \ion{Ca}{19} lines include the resonance line (line $w$, 3.17735~\AA: notation of \cite{gab72}), intercombination lines ($x$,  3.18941~\AA; $y$, 3.19291~\AA), and forbidden line ($z$, 3.21095~\AA) (quoted wavelengths are from SOLFLEX measurements: \cite{see89}). Prominent dielectronic satellites emitted by Li-like Ca (\ion{Ca}{18}) occur near the \ion{Ca}{19} lines, with wavelengths mostly between lines $w$ and $z$. Other dielectronic satellites, emitted by Li-like Ca (\ion{Ca}{17}), have been observed in solar flare spectra by SOLFLEX as well as from high-temperature plasmas in the Alcator C-Mod tokamak device \citep{rice14,rice15,rice17}.

Here we discuss a very large ({\em GOES}\/ class X5.3) flare on 2001 August~25 observed by the DIOGENESS scanning crystal spectrometer on the {\em CORONAS-F}\/ spacecraft. Scans were made over the complete duration of the flare with two of the four channels covering the wavelength region of the \ion{Ca}{19} lines and \ion{Ca}{18} satellites. The spectra were averaged over five time periods each with similar temperature, defined by $T_{\rm GOES}$ derived from the emission ratio of the two channels of {\em GOES}\/; details are described in Section~\ref{sec:Diogeness_sp}. During the flare decay stage, when the temperature was lower, lines of lower ionization stages are also evident as are lines of H-like and He-like argon (\ion{Ar}{18}, \ion{Ar}{17}) and associated dielectronic satellites. We compare the observed spectra with isothermal synthetic spectra over the 3.05~~--~3.35~\AA\ range generated by a purpose-written code, developed for this work and described in Section~\ref{sec:Synth_Sp}, based on various published sources and new calculations of calcium and argon satellite lines made with the Cowan Hartree--Fock atomic code \citep{cow81}. A grid of spectra with a large range of input electron temperatures $T_e$ allowed this comparison to be made. The averaged DIOGENESS spectra over the five temperature intervals were compared with the synthetic spectra calculated for temperature equal to $T_{\rm GOES}$ (Section~\ref{sec:obs_calc_sp}). Most of the spectral features are reproduced with high accuracy although the lower-temperature \ion{Ca}{17} lines were found to be more intense than calculated. This is attributed to uncertainties in the very small Ca$^{+16}$ ion fractions at these temperatures; increasing these then gives much improved agreement. Justification for this adjustment is provided by trial calculations in which the effect of 10\% uncertainties in the ionization and recombination rates on the ion fractions are examined (paper in preparation).

A number of spectral line features -- dielectronic satellites of Ca and Ar ions -- not previously noted are observed and identified in the DIOGENESS spectra. The synthetic spectra are also compared with published spectra from SOLFLEX during solar flares and with highly ionized calcium spectra obtained from the Alcator C-Mod tokamak device. The present analysis will be applied to solar flare spectra in the {\em SMM}\/ data archive and to spectra expected from the crystal spectrometer ChemiX \citep{sia16}, which is part of the payload of the Sun-orbiting {\em Interhelioprobe}\/ spacecraft, to be launched in 2025 or 2026.

% Section 2
\section{DIOGENESS SPECTRA} \label{sec:Diogeness_sp}

% Subsection 2.1
\subsection{DIOGENESS Instrument and Calibration} \label{sec:instr_calib}

The {\em CORONAS-F}\/ spacecraft was operational between 2001 and 2006, and had a near-polar orbit with period 94.9~minutes, with up to 35-minute spacecraft night periods and occasional interruptions due to passages through enhanced particle radiation associated with the South Atlantic Anomaly and auroral oval regions near each pole. The bent crystal spectrometer RESIK (REntgenovsky Spektrometr s Izognutymi Kristalami) built by the Space Research Centre (SRC) group obtained solar flare spectra in the 3.36~--~6.05~\AA\ range which have been extensively used for element abundance determinations and other investigations (see \cite{jsyl05,jsyl12} and references therein). The SRC companion DIOGENESS instrument operated for only a few weeks at the beginning of the spacecraft mission, 2001 August~16 to September~17, but in its short lifetime it observed several large flares (for flare list and instrumental details, see \cite{jsyl15}) including an X5.3 flare on 2001 August~25 (SOL2001-08-25T16:45), one of the largest in Solar Cycle~23.  DIOGENESS consisted of four quartz crystals mounted on a single rotatable shaft that were rocked back and forth during the observations. Thus, X-rays from a solar flare were incident at various angles $\theta$, giving diffraction at wavelengths $\lambda$ according to Bragg's law $\lambda = 2d \,{\rm sin}\,\theta$ for first-order diffraction ($d$ is the crystal lattice spacing and is equal to 6.6855~\AA\ for the quartz crystals used). The diffracted radiation was detected by double proportional counters, with calibration sources illuminating the rear section of each detector. This type of double proportional counter arrangement was previously used for X-ray detection in solar instruments aboard the {\em Prognoz} and {\em Interball-tail} satellites constructed by the SRC group. Channels~2 and 3 of DIOGENESS covered narrow wavelength bands around the resonance lines of \ion{S}{15} (5.04~\AA) and \ion{Si}{13} (6.65~\AA) respectively. The August~25 flare spectra discussed here were obtained by channels~1 and 4, including lines of He-like Ca (\ion{Ca}{19}) with nearby \ion{Ca}{18} dielectronic satellite lines (3.17~--~3.21~\AA); the wavelength ranges were 3.12~--~3.32~\AA\ (channel~1) and 3.04~--~3.24~\AA\ (channel~4), so the overlapping region (3.12 -- 3.24~\AA) included the \ion{Ca}{19} lines. In some spectra, an \ion{Fe}{25} emission line in second order diffraction, at 3.1463~\AA, is apparent. Channels 1 and 4 were arranged in a ``Dopplerometer'' mode, i.e., with the two crystals (of quartz $10\bar11)$ facing each other such that spatial displacements of the flare emitting region could be distinguished from spectral shifts due to Doppler motions.

An absolute intensity calibration was first established, with the instrument's effective area $A_{\rm eff}$ (cm$^2$) given by

% Eq. 1
\begin{equation}
A_{\rm eff} = A_{\rm window} \times \kappa_{\rm window} \times E_{\rm det} \times R_{\rm int}\,{\rm ,}
\end{equation}

\noindent where $A_{\rm window}$ is the detector window area, $\kappa_{\rm window}$ and $E_{\rm det}$ are the absorption coefficients of the beryllium window and detector gas respectively, and $R_{\rm int}$ the integrated reflectivity of the crystal. Although no pre-launch measurements were made of the detectors, data are available from measurements made on identical flight spares, giving for the \ion{Ca}{19} line $w$ wavelength (3.176~\AA) $\kappa_{\rm window}= 0.78$, $A_{\rm window} = 0.2854 \pm 0.006$~cm$^2$, and $E_{\rm det} = 0.57$. An on-line X-ray optics toolkit (XOP: \cite{xop04a}) was used to evaluate $R_{\rm int}$, which was found to be $3.4 \times 10^{-5}$ at 3.17~\AA\ for quartz $10\bar11$ crystal (a re-measured value of a flight spare crystal gave $4.0 \times 10^{-5}$). The resulting value of $A_{\rm eff}$ at 3.17~\AA\ was determined to be $4.31 \times 10^{-6}$~cm$^2$, with an estimated uncertainty of 20\%. The rotational scanning speed was 56~arcmin~s$^{-1}$ and data gathering interval 0.40125~s.

% Subsection 2.2
\subsection{DIOGENESS Spectra} \label{sec:DIOG_sp}

\begin{figure}
\epsscale{0.9}
\plotone{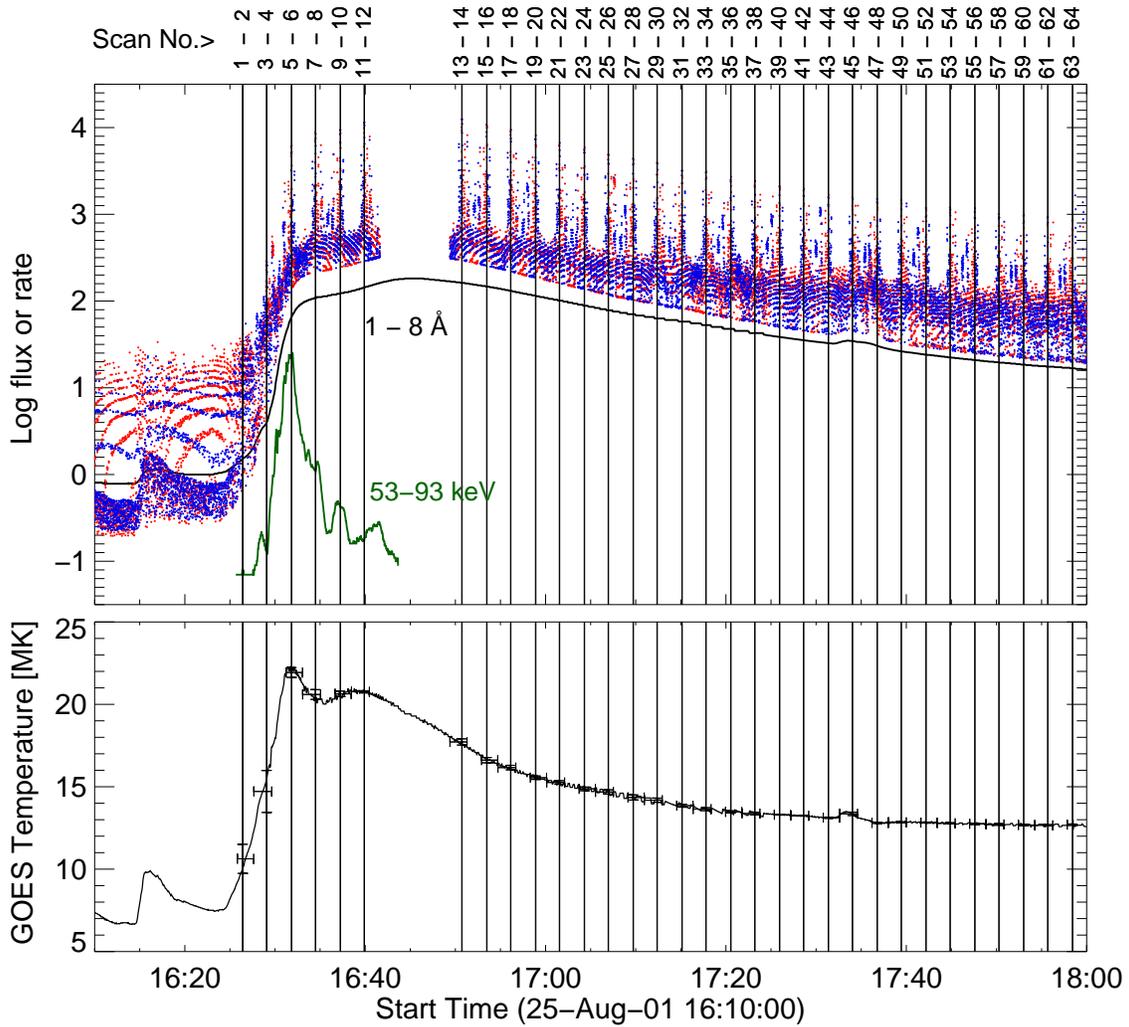}
%\vspace{25mm}
\caption{Upper panel: Logarithmic plots of {\em GOES}\/ light curve in the 1--8~\AA\ channel, DIOGENESS channel~1 (red dots) and channel~4 (blue dots), with {\em Yohkoh}\/ Hard X-ray Telescope (53--93~keV) (green curve) photon count rates during the 2001 August~25 X5 flare are shown. The DIOGENESS points show peaks due to the \ion{Ca}{19} line group as the crystals repeatedly scanned over their ranges. The data gap between 16:42~UT and 16:49~UT is due to telemetry loss. Lower panel: Temperature derived from the intensity ratio of the two {\em GOES}\/ channels (horizontal error bars are periods over which averages were obtained, vertical error bars are standard deviations in temperature estimates).  \label{fig:DIOG_1-4_GOES_time} }
\end{figure}

Figure~\ref{fig:DIOG_1-4_GOES_time} (top panel) shows the time variations in the photon count rate of DIOGENESS channels~1 and channel~4, and the {\em GOES}\/ $1-8$~\AA\ irradiance (black curve) during the August~25 flare over the period 16:09~--~17:50~UT. A small precursor apparent in both the DIOGENESS and {\em GOES}\/ data occurred at 16:17~UT, followed by a sharp rise (16:26~--~16:34~UT) to maximum which occurred at about 16:45~UT. The impulsive stage during this sharp rise is marked by hard X-ray ($53 - 93$~keV) emission observed by the {\em Yohkoh}\/ Hard X-ray Telescope. The flare decay, with a short minor burst at 17:35~UT, was observed by DIOGENESS until 17:50~UT. A gap in the DIOGENESS data at 16:42~--~16:48~UT is due to a brief loss of telemetry. The DIOGENESS data points show peaks corresponding to the \ion{Ca}{19} spectral line group as the spectrometer drive scanned back and forth over its range in a period of $\sim 140$~s. Some 64 channel~1 and 4 spectra were obtained. The lower panel of this figure shows the temperature $T_{\rm GOES}$ derived from the ratio of emission in the two {\em GOES}\/ channels averaged over short time intervals (indicated by horizontal error bars) with standard deviations (vertical error bars).  {\em Yohkoh}\/ HXT and Soft X-ray Telescope (SXT) images of the flare, which was located in the south--east part of the solar disk (S17 E34), were shown by \cite{jsyl15}. The fast-changing morphology of the sources during the initial phase of this flare is discussed by \cite{sia04}.

\begin{figure}
\epsscale{0.9}
\plotone{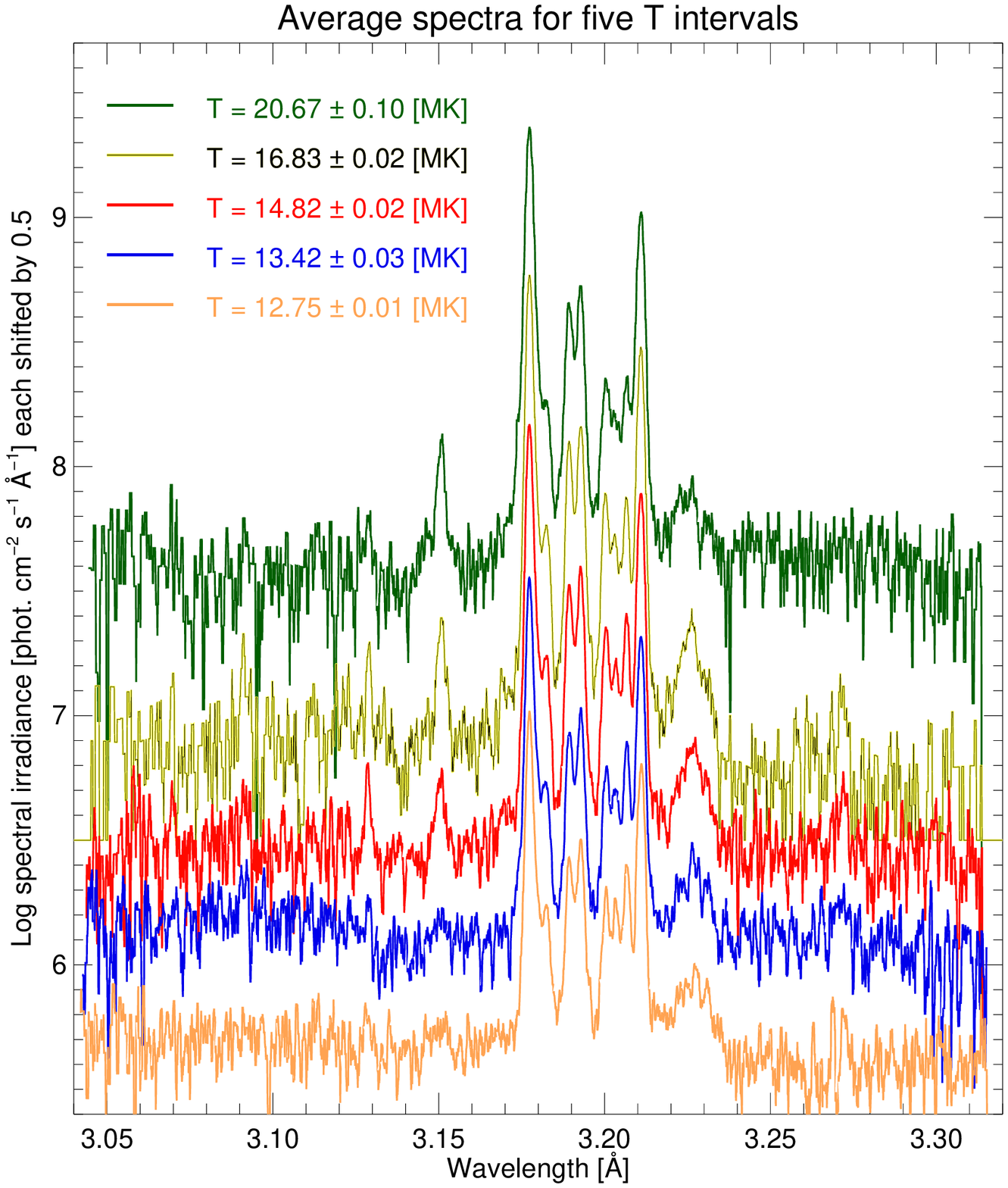}
%\vspace{30mm}
\caption{DIOGENESS spectra plotted on a logarithmic scale during the 2001 August~25 flare averaged over five time intervals defined by $T_{\rm GOES}$\/ indicated in the figure legend. The vertical scale units are shown, valid for the $T_{\rm GOES} = 12.8$~MK spectrum, with successive higher-temperature spectra increased by 0.5 in the logarithm for clarity. See Table~\ref{tab:line_wavelengths} for identification of principal spectral features.  \label{fig:DIOG_av_sp_five_temps}}
\end{figure}

As some of the 64 DIOGENESS spectra have low or medium statistical significance, channel~1 and 4 spectra were combined and averaged over time periods that correspond to five intervals when the derived $T_{\rm GOES}$ (in MK) was in the ranges 12--13, 13--14, 14--16, 16--18, and 18--21, with average values (MK) of 12.8, 13.4, 14.8, 16.8, and 20.7. They are shown as logarithmic plots in Figure~\ref{fig:DIOG_av_sp_five_temps} in spectral irradiance units (photon cm$^{-2}$~s$^{-1}$~\AA$^{-1}$) using the calibration data of Section~\ref{sec:instr_calib} (spectra for $T_{\rm GOES} \geqslant 12.8$~MK are increased by 0.5 in the logarithm for clarity); the spectral irradiances were divided by the emission measure $EM_{\rm GOES}$ derived from $T_{\rm GOES}$. The five intervals exclude the initial flare rise since the plasma conditions could be non-stationary and non-equilibrium. The background in each of the spectra shown is due to fluorescence and secondary emissions related to magnetospheric high-energy particles; analysis of the spectra included this background to preserve as far as possible the integrity of low-intensity spectral features discussed further in this work. The spectral range of channel~1 includes not only the \ion{Ca}{19} lines and \ion{Ca}{18} satellites but also satellites due to lower ionization stages on the long-wavelength side of \ion{Ca}{19} line $z$. The short-wavelength part of the channel~4 range includes \ion{Ar}{17} ($1s^2 - 1snp$, $n=5, 6, 7, 8$: called here $w5$, $w6$, $w7$, and $w8$) and \ion{Ar}{18} (Ly$\beta$) lines. Dielectronic satellites associated with these lines are also evident, as we shall discuss later.

%FFFFFFFFFFFFFFFFFFFFFFFFFFFFFFFFFFFFFFFFFFFFFFFFFFFFFFFFFFFFFFFFFFFFFFFFFFFFFFFFFFFFFFFFFFFFFFFFFFFFFFFFFFFFFFFFFFFFFFFFFFFFFFFFFFFFFFFFFFFF
% Figure 1 (DIOG count rates, HXT, GOES light curve + GOES T

%FFFFFFFFFFFFFFFFFFFFFFFFFFFFFFFFFFFFFFFFFFFFFFFFFFFFFFFFFFFFFFFFFFFFFFFFFFFFFFFFFFFFFFFFFFFFFFFFFFFFFFFFFFFFFFFFFFFFFFFFFFFFFFFFFFFFFFFFFFFF

% Figure 2: DIOGENESS spectra  averaged over five temperature intervals.
%FFFFFFFFFFFFFFFFFFFFFFFFFFFFFFFFFFFFFFFFFFFFFFFFFFFFFFFFFFFFFFFFFFFFFFFFFFFFFFFFFFFFFFFFFFFFFFFFFFFFFFFFFFFFFFFFFFFFFFFFFFFFFFFFFFFFFFFFFFFF

% Section 3
\section{SYNTHETIC SPECTRA}\label{sec:Synth_Sp}

% Subsection 3.1
\subsection{Calculation}\label{sec:Calc}

The {\sc chianti} (v.~8) atomic database and code \citep{der97,lan06} allow the calculation of \ion{Ca}{19} spectra and \ion{Ca}{18} dielectronic satellites, although refinements to the \ion{Ca}{19} line intensities and also satellites from \ion{Ca}{17} and lower ionization stages are not included. These satellites are important in the August~25 flare spectra, so for the purpose of this study we chose to synthesize spectra using a specially written program in Interactive Data Language (IDL), incorporating atomic data from various sources. The calculated spectra include the principal \ion{Ca}{19} lines, which dominate the spectrum, and dielectronic satellites of \ion{Ca}{18}, \ion{Ca}{17},  \ion{Ca}{16}, and \ion{Ca}{15}. These satellites are mostly due to $1s - 2p$ transitions in the presence of spectator (non-participating) electrons. Lines of He-like Ar (\ion{Ar}{17}) lines ($w4$, $w5$, $w6$, and $w7$) and the \ion{Ar}{18} $1s - 3p$ (Ly$\beta$) line were included, as were associated \ion{Ar}{16} and \ion{Ar}{17} dielectronic satellites.

A grid of synthetic spectra in the range 3.00~--~3.35~\AA\ based on calculated line and continuum intensities was constructed for temperatures log~$T_e = 6.48 - 8.0$ ($T_e = 3 - 100$~MK) in 0.02 dex steps and further interpolated in steps of 0.1~MK. Spectra in this grid were compared with observed spectra during the August~25 flare for deducing temperatures on an isothermal assumption. A value of emission measure, $N_e^2\,V$ ($N_e = $ electron density, $V = $ emitting volume), was derived from the factor that the synthetic spectrum in spectral irradiance units is multiplied by in order to agree with the observed spectrum. An isothermal assumption is valid for the observed spectra shown in Figure~\ref{fig:DIOG_av_sp_five_temps}, which exclude those in the initial sharp rise in emission. Ionization equilibrium was assumed as is justified for flare plasma densities except when the temperature is rapidly changing. Ion fractions as a function of temperature were taken from {\sc chianti} v.~8 \citep{delz15} (these are very close to those of \cite{bry09}). From observations with the {\em SMM}\/ Bent Crystal Spectrometer \citep{jsyl98}, the Ca abundance is known to vary from flare to flare; the average value, $A({\rm Ca}) = 6.76$ (log scale with $A({\rm H}) = 12$), was taken for the calculated spectra and for the observed August~25 spectra. For the argon abundance, we took $A({\rm Ar}) = 6.45$ based on an extensive analysis of RESIK flare observations \citep{jsyl10b}. Thus the Ar/Ca abundance ratio is 0.33, which is less than the photospheric ratio (0.87: \cite{asp09}), but is accounted for by an expected factor-of-three or more enhancement of elements with low ($\lesssim 10$~eV) first ionization potentials (FIPs) including calcium but not argon (see \cite{phi08}, chapter~11).

% Subsection 3.2
\subsection{Ca XIX Lines}\label{sec:CaXIX}

The principal lines of He-like Ca (\ion{Ca}{19}) are $w$ ($1s^2\,^1S_0 - 1s2p\,^1P_1$), $x$ ($1s^2\,^1S_0 - 1s2p\,^3P_2$), $y$ ($1s^2\,^1S_0 - 1s2p\,^3P_1$), and $z$ ($1s^2\,^1S_0 - 1s2p\,^3S_1$). The excitation of these lines is principally by electron collisions; here we used excitation rates from the distorted wave calculations of \cite{bel82b}. These include excitation by recombination from H-like Ca, important at relatively high temperatures (e.g., $\sim 21$\% contribution to line~$z$ at 20~MK), and ionization of Li-like Ca, important at the few per cent level for line~$z$ at relatively low ($\lesssim 5$~MK) temperatures. We used the wavelengths of \ion{Ca}{19} lines from \cite{see89}, measured from SOLFLEX spectra.

Since the \cite{bel82b} work on \ion{Ca}{19} collision strength calculations, more detailed close-coupling collision strengths have become available, so comparison with these should be made. The close-coupling calculations of \cite{whiteford01} are those used in the {\sc chianti} software package for the \ion{Ca}{19} lines, and include radiation damping of the auto-ionizing resonances that are more important for the excitation of \ion{Ca}{19} lines $x$, $y$, and $z$. The more recent work of \cite{agg12} shows some differences from \cite{whiteford01} although these are only important for ultraviolet transitions. A comparison of the \cite{bel82b} collision strengths for energies between 400~Ry and 800~Ry and those of \cite{agg12} shows differences that are extremely small, of order 1\%, for excitation to the upper level of line $w$, and only a few percent for other transitions. The largest difference (up to 6\%) is for excitation to the upper level of line $y$. Thus, use of the \cite{bel82b} collision strength data rather than the close-coupling data should result in negligible differences to calculated spectra. 

% Subsection 3.3
\subsection{\ion{Ca}{18} -- \ion{Ca}{15} Dielectronic Satellites}

The theory of the formation of dielectronic satellite lines is given by \cite{gab72} and applied to \ion{Ca}{18} satellites by \cite{bel82b}, so we give only brief details here. A \ion{Ca}{18} satellite line is formed by the dielectronic capture of a free electron by the He-like ion Ca$^{+18}$ resulting in a doubly excited state of the Li-like ion Ca$^{+17}$:

% Eq. 2
\begin{equation}
{\rm Ca}^{+18}(1s^2) + e^{-} \leftrightarrows {\rm Ca}^{+17}(1s\,2p\,nl) \rightarrow {\rm Ca}^{+17} (1s^2\,nl).
\label{eq:diel_exc}
\end{equation}

\noindent The doubly excited state may de-excite either radiatively or by auto-ionization to Ca$^{+18}$ and a free electron (hence the double arrow in Equation~\ref{eq:diel_exc}). If there is radiative de-excitation, satellite lines in the transition array $1s^2\,nl - 1s\,2p\,nl$ result. The principal $n=2 - 4$ dielectronic lines cover the wavelength range 3.175~--~3.269~\AA; for increasing $n$, the satellites converge on the \ion{Ca}{19} lines.

Following \cite{gab72}, the satellite line irradiance at the distance of the Earth ($I_{\rm s}$: photon cm$^{-2}$~s$^{-1}$) is

%Eq. 3
\begin{equation}
I_{\rm s} = \frac { N_e\, N({\rm Ca}^{+18}) \, V}{4 \pi\, ({\textsc{au}})^2} \,\, \frac {2.06\times
10^{-16}\, F_{\rm s}\,{\rm exp} \,(-E_s /k_B\, T_e)} {g_1 T_e^{3/2}}\qquad\ {\rm [photons}\,\,{\rm
cm}^{-2} \,\,{\rm s}^{-1}{\rm ]}, \label{eq:sat_int}
\end{equation}

\noindent where $E_{\rm s}$ is the excitation energy of the line's upper state above the ground state of the He-like ion and $g_1$ the statistical weight of the ground level ($1s^2\,^1S_0$) of Ca$^{+18}$ (so $g_1 = 1$). $k_B$ is Boltzmann's constant, and $N({\rm Ca}^{+19})$ is the number density of He-like Ca ions. The satellite intensity factor $F_{\rm s}$ is given by

% Eq. 4
\begin{equation}
F_{\rm s} = \frac { g_s A^{\rm r}\, A^{\rm a}} {A^{\rm a} + \Sigma A^{\rm r}},
\label{eq:Fofs}
\end{equation}

\noindent where $A^{\rm r}$ and $A^{\rm a}$ are transition probabilities from the satellite's upper state by radiation and autoionization respectively and $g_s$ the statistical weight of the upper level of the satellite line transition. The summation is of radiative transition probabilities from the upper state to all possible lower states.

Values of $F_{\rm s}$ and wavelengths for the \ion{Ca}{18} satellites are available from \cite{bel82b} and a number of other works (e.g., \cite{vai78}). Here, for consistency with calculations of satellites of lower ionization stages, apart from a few exceptions we used the Cowan Hartree--Fock atomic code \citep{cow81} to calculate the necessary data. This code was run previously in studies we made of X-ray dielectronic satellites of Li-like K and Cl seen with the RESIK instrument \citep{jsyl10a,bsyl11}. The code\footnote{The Cowan atomic program package is currently hosted at Trinity College Dublin: https://www.tcd.ie/Physics/people/Cormac.McGuinness/Cowan/}, outlined in \cite{cow81} (Chapters~8 and 16) and described in more detail by \cite{mer76}, calculates energy levels and radiative and autoionization probabilities using Slater--Condon theory with pseudo-relativistic corrections. It has been adapted for personal computers (A.~Kramida, private communication, 2014) which was how the code was run for this work. The code accepts input for a given satellite line array with scale factors for Slater parameters, specified in recent updates to the program; the output includes values for excitation energies, wavelengths, oscillator strengths, transition probabilities, and $F_{\rm s}$ for optically allowed transitions. This restriction excludes the  \ion{Ca}{18} satellites $o$ and $p$ (3.2688~\AA, 3.2636~\AA) for which the transitions are optically forbidden; for these we used the data of \cite{bel82b}. The Cowan calculations were carried out for \ion{Ca}{18} satellite arrays $1s^2\,nl - 1s2p\,nl$ with $n$ up to 6 and all possible $l$ values. The absolute wavelengths are stated by \cite{mer76} to be accurate to about 0.2\% for transitions considered here, i.e., about 6~m\AA. Our calculations of absolute wavelengths indicate a higher precision is achieved; the Cowan wavelengths are generally less than both those of \cite{see89} measured from {\em P78-1} SOLFLEX spectra and those of \cite{rice14} measured from Alcator C-Mod tokamak spectra by an amount $\Delta \lambda$ between 2~m\AA\ and 4~m\AA, with an average value of $\Delta \lambda = 3.2$~m\AA. This amount was added to the Cowan wavelengths in the synthesis code. It was checked with runs of the Cowan code for satellites with very high $n$ which converge on the \ion{Ca}{19} lines $w$, $x$, and $y$, giving $\Delta \lambda \approx 3$~m\AA. An exception is the medium-importance satellite $m$ for which the Cowan wavelength (3.1926~\AA) is more than the calculated wavelengths of \cite{bel82b} (3.1880~\AA) and \cite{vai78} (3.1890~\AA) by about 4~m\AA. Laser plasma X-ray wavelength measurements by \cite{fel74} for various He-like ion spectra show that the wavelength is near that of \cite{bel82b}, and accordingly this wavelength was adopted here. Note that in all solar and Alcator spectra, the highest-intensity satellite $j$ is indistinguishable from \ion{Ca}{19} line~$z$.

Satellites of \ion{Ca}{17}, \ion{Ca}{16}, and \ion{Ca}{15} were similarly calculated with the Cowan code. These satellites occur to the long-wavelength side of \ion{Ca}{19} line $z$ ($\gtrsim 3.21$~\AA). Their intensities are given by Equations~(\ref{eq:sat_int}) and (\ref{eq:Fofs}) with $g_1 = 2$ for \ion{Ca}{17} satellites (excitation from the ground level $1s^2\,2s\,\,^2S_{1/2}$ of Ca$^{+17}$ with statistical weight 2), 1 for \ion{Ca}{16} (excitation from $1s^2\,2s^2\,\,^1S_0$ of Ca$^{+16}$), and 2 for \ion{Ca}{15} (excitation from $1s^2\,2s^2\,2p\,\,^2P_{1/2}$ of Ca$^{+15}$). Several \ion{Ca}{17} satellites are particularly prominent in the DIOGENESS solar spectra as well as in the flare spectra of \cite{see89} and the Alcator spectra of \cite{rice14}; we took the measured wavelengths of these from \cite{see89}. (Lines with identical transitions are also seen in solar flare \ion{Fe}{23} spectra: \cite{lem84}.) The Cowan wavelengths for the same lines are smaller by $\sim 3$~m\AA, i.e., a similar amount to that of the \ion{Ca}{18} satellites, so for the remaining satellites, 3.2~m\AA\ was added to the Cowan wavelengths.

Table~\ref{tab:line_wavelengths} gives data for important lines of calcium ions included in the spectral synthesis program, including line notation and transitions, wavelengths adopted and measured (\cite{see89,rice14}). The line notation is from \cite{gab72} for the \ion{Ca}{19} and \ion{Ca}{18} lines, and from \cite{see89} for the \ion{Ca}{17} satellite features seen in SOLFLEX spectra. Table~\ref{tab:line_wavelengths} also gives, for the dielectronic satellites, values of $F_{\rm s}$ from the Cowan program and for comparison values from either \cite{vai78} or \cite{saf79}. Approximately 2000 other weaker satellites of calcium ions are included in the program. There is $\sim 10$\% agreement between the Cowan $F_{\rm s}$ values and those of \cite{vai78} or \cite{saf79} for the more intense satellites ($F_{\rm s} \gtrsim 10^{14}$~s$^{-1}$); for weaker satellites, a factor-of-two agreement is more typical. Such errors should not affect the general appearance of the spectrum. Note that neither this table, nor our calculations, includes the \ion{Fe}{25} line (transition $1s^2\,^1S_0 - 1s3p\,^1P_1$) with wavelength 1.5731~\AA\ (observed in second diffraction order at 3.1463~\AA) which is apparent in high-temperature DIOGENESS spectra.

Inner-shell excitation is significant for several \ion{Ca}{18} satellites, notably $o - v$. For these satellites, we used the collisional excitation rates calculated by \cite{bel82b}. The excitation of the \ion{Ca}{17} $\beta$ line (transition $1s^2\,2s^2\,^1S_0 - 1s\,2s^2\,2p\,^1P_1$) is also significant, but no published data for the collision excitation rates currently exist. However, temperature-averaged collision strengths for this and the $1s^2\,2s^2\,^1S_0 - 1s\,2s^2\,2p\,\,^3P_1$ transition have been calculated by \cite{man83} for \ion{Fe}{23}, the equivalent Fe ion. $Z$-scaling laws  ($Z = $ atomic number) given by \cite{bur92} were used to calculate temperature-averaged collision strengths for \ion{Ca}{17} line $\beta$, from which excitation rates were derived. For a broad temperature range, the dielectronic and inner-shell contributions to this line are comparable. Although the $F_{\rm s}$ value used here differs from that of \cite{saf79} (Table~\ref{tab:line_wavelengths}), there is little effect on the total intensity of this line. Inner-shell contributions to some other \ion{Ca}{16} and \ion{Ca}{15} lines are possible but are much smaller and have been neglected.

\begin{figure}
\epsscale{0.9}
\plotone{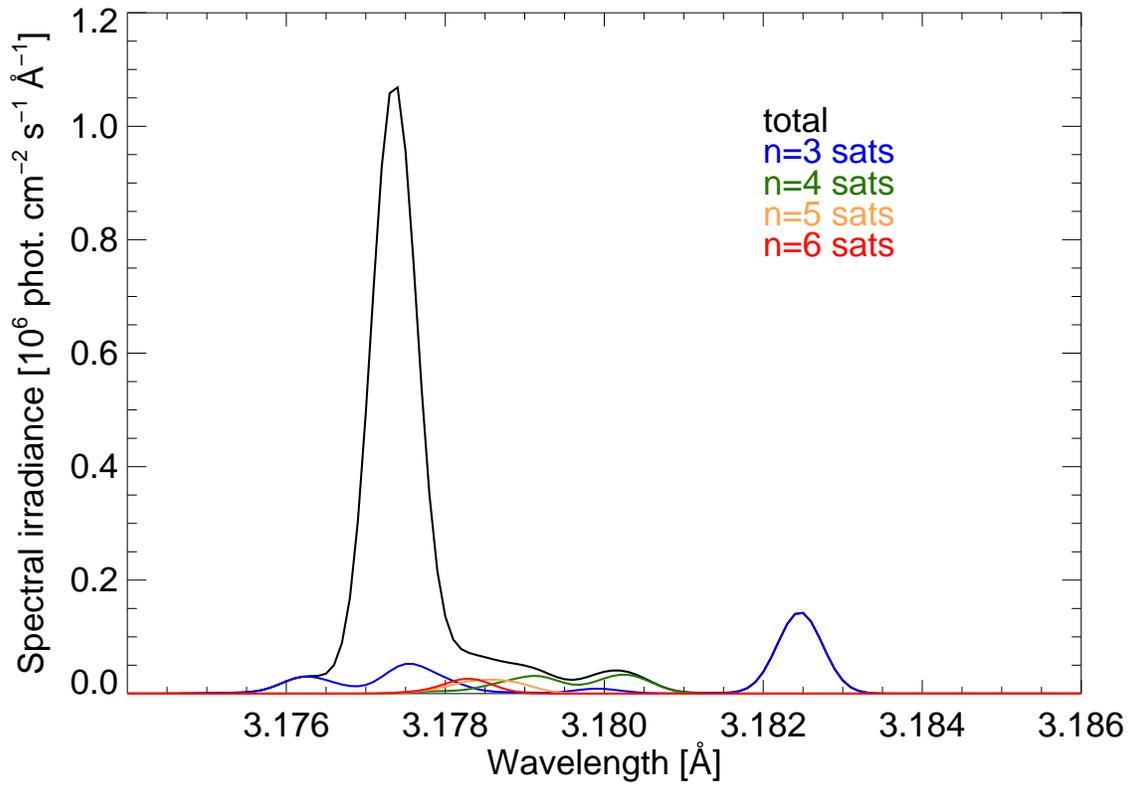}
%\vspace{25mm}
\caption{Convergence of high-$n$ dielectronic satellites on to the Ca XIX $w$ line: theoretical spectrum for $T = 13.8$~MK. Each satellite group is color-coded, and the key indicated in the legend.   \label{fig:Ca_high_n_sats_conv}}
\end{figure}

The convergence of the high-$n$ satellites calculated here is illustrated by Figure~\ref{fig:Ca_high_n_sats_conv}, showing the summed contribution of satellites with $n=3$, 4, 5, and 6 near \ion{Ca}{19} line $w$ calculated for $T = 13.8$~MK. For this temperature, we estimate that the total emission within 1.15~m\AA\ of the central wavelength of  \ion{Ca}{19} line~$w$, 8.6\% is due to unresolved $n=3$ satellites, 1.4\% to $n=4$ satellites, 2.3\% to $n=5$ satellites, and 2.1\% to $n=6$ satellites. Using the data of \cite{bel82b}, the contribution of $n>7$ satellites is approximately 1.6\%.

%FFFFFFFFFFFFFFFFFFFFFFFFFFFFFFFFFFFFFFFFFFFFFFFFFFFFFFFFFFFFFFFFFFFFFFFFFFFFFFFFFFFFFFFFFFFFFFFFFFFFFFFFFFFFFFFFFFFFFFFFFFFFFFFFFFFFFFFFFFFF
% Figure 3: Convergence of high-n sats on to Ca XIX w line
%FFFFFFFFFFFFFFFFFFFFFFFFFFFFFFFFFFFFFFFFFFFFFFFFFFFFFFFFFFFFFFFFFFFFFFFFFFFFFFFFFFFFFFFFFFFFFFFFFFFFFFFFFFFFFFFFFFFFFFFFFFFFFFFFFFFFFFFFFFFF
% Figure 4: Synthetic spectrum showing low ioniz. stages of Ca.
\begin{figure*}
\epsscale{0.9}
\plotone{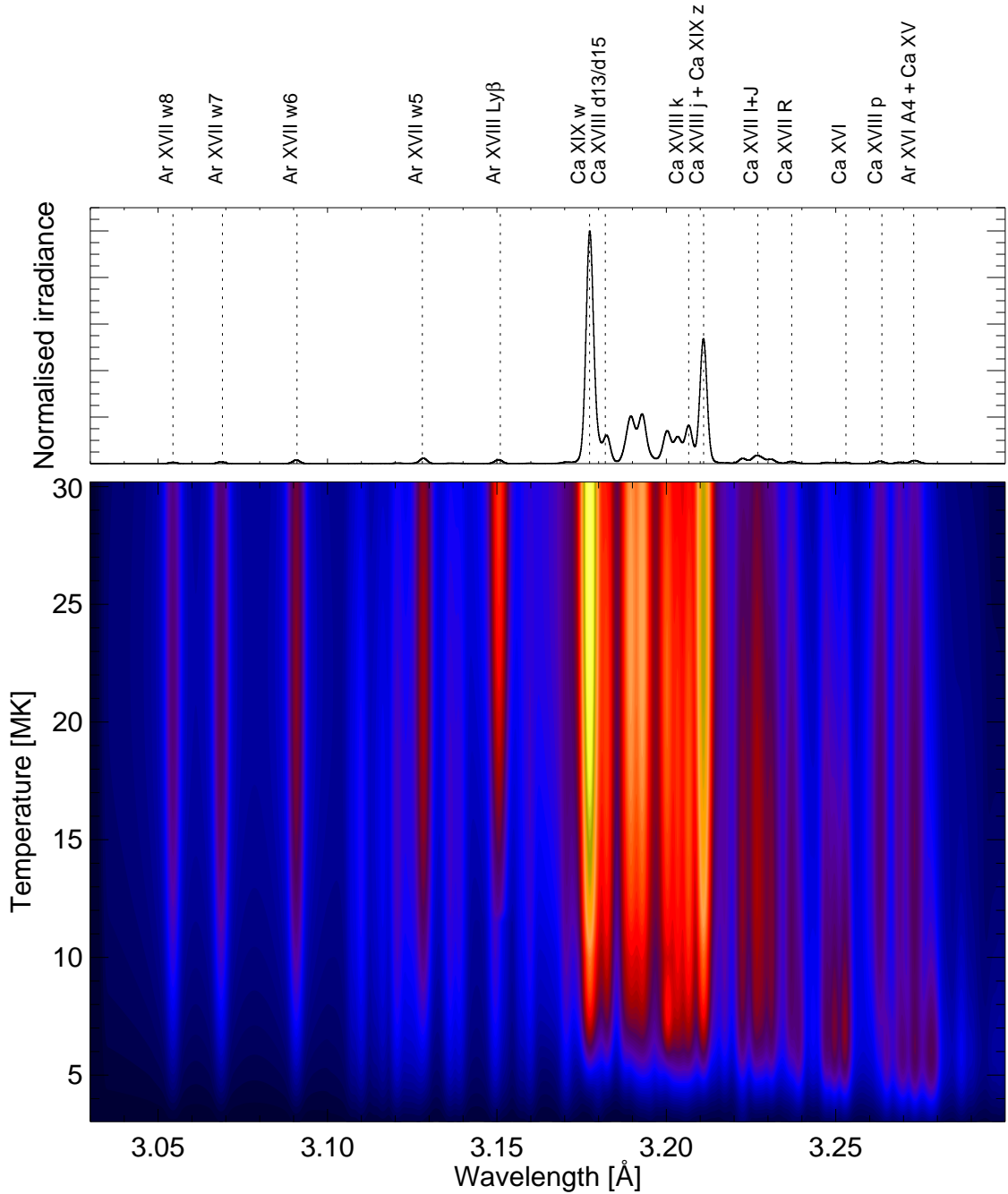}
%\vspace{25mm}
\caption{Synthetic spectra in the range $3.00 - 3.30$~\AA\ for temperatures in the range 4~MK (bottom) to 30~MK (top), color-coded with blue, orange, red, yellow indicating increasing intensities. The chief lines are identified at the top of the figure with a theoretical spectrum with temperature equal to 13.8~MK. (For line widths, see text.)    \label{fig:theoretical_sp}}
\end{figure*}
%FFFFFFFFFFFFFFFFFFFFFFFFFFFFFFFFFFFFFFFFFFFFFFFFFFFFFFFFFFFFFFFFFFFFFFFFFFFFFFFFFFFFFFFFFFFFFFFFFFFFFFFFFFFFFFFFFFFFFFFFFFFFFFFFFFFFFFFFFFFF

%TTTTTTTTTTTTTTTTTTTTTTTTTTTTTTTTTTTTTTTTTTTTTTTTTTTTTTTTTTTTTTTTTTTTTTTTTTTTTTTTTTTTTTTTTTTTTTTTTTTTTTTTTTTTTTTTTTTTTTTTTTTTTTTTTTTTTTTTTTTT
% Table 1: Wavelengths of principal lines in the 3.00 - 3.30 A range.
% See Notes 4.2.16, 3.2.17, 25-29.1.18. (Jan. 2018: Combined all tables in earlier versions of text. Line nos. in IDL program, 22.3.18)
\begin{deluxetable}{llllllll}
\tabletypesize{\tiny}
\tablecaption{Principal Lines in the range 3.00--3.30~\AA. \label{tab:line_wavelengths}}
\tablewidth{0pt}
\tablehead{
\colhead{Ion} & \colhead{Transition} & \colhead{Key$^a$} &  \colhead{Wavel.}     &\colhead{Wavel.}        & \colhead{$F(s)^d$}  &\colhead{$F(s)^e$} &\colhead{$F(s)^f$} \\
              &                      &                   &\colhead{(Adopted)$^b$}&\colhead{(Measured)$^c$}& \colhead{(s$^{-1}$)}&\colhead{(s$^{-1}$)}&\colhead{(s$^{-1}$)}\\}
\startdata

Ar XVII  & $1s^2\,^1S_{0} - 1s\,8p\,^1P_{1}$                  & $w8$   & 3.0544  & 3.0544   &        &            \\ %2057
Ar XVII  & $1s^2\,^1S_{0} - 1s\,7p\,^1P_{1}$                  & $w7$   & 3.0686  & 3.0686   &        &            \\ %2056
Ar XVII  & $1s^2\,^1S_{0} - 1s\,6p\,^1P_{1}$                  & $w6$   & 3.0908  & 3.0908   &        &            \\ %2055
Ar XVII  & $1s^2\,^1S_{0} - 1s\,5p\,^1P_{1}$                  & $w5$   & 3.12834 & 3.12834                        \\%2054
Ar XVI   & $1s^2\,2p\,^2P - 1s 2p 8p\,(^3P)\,^2D $            & A8     & 3.1361  & 3.135    & 5.12(12) & 4.18(12) \\ %2215,2220
Ar XVI   & $1s^2\,2p\,^2P - 1s 2p 7p\,(^3P)\,^2D $            & A7     & 3.1495  & 3.146    & 7.89(12) & 6.13(12) \\ %2186,2190
Ar XVIII & $1s\,^2S_{1/2} - 2p\,^2P_{1/2,3/2}$                & Ly$\beta$ & 3.15062 & 3.15062                     \\%2058
Ar XVI   & $1s^2\,2p\,^2P - 1s 2p 6p\,(^3P)\,^2D$             & A6     & 3.1704  & 3.169    & 1.31(13) & 1.08(13) \\ %2151,2154
Ca XIX   & $1s^2\,^1S_{0} - 1s\,2p\,^1P_{1}$                  & $w$    & 3.17735 & 3.17735  & 3.1773  \\%0
Ca XVIII & $1s^2\,4p\,^2P - 1s\,2p\,4p\,^2D$                  & $n=4$  & 3.1802  & 3.17926  & 9.89(13) &        \\
Ca XVIII & $1s^2\,3p\,^2P - 1s\,2p\,3p\,(^1P)\,^2D$& $d15$,$d13$ &3.1818 &3.1824 & 3.14(14) &          & 5.55(14) \\%54,56
%Ca XVIII & $1s^2\,3p\,^2P_{3/2} - 1s\,2p\,3p\,(^1P)\,^2D_{5/2}$& $d13$ & 3.1823  & 3.1824   & 1.81(14) &          & 3.84(14) \\%54
Ca XVIII & $1s^2\,2p\,^2P_{3/2} - 1s\,2p^2\,^2S_{1/2}$        & $m$    & 3.1880  &          & 7.03(13) & 4.10(13) & 2.99(13) \\%16
Ca XIX   & $1s^2\,^1S_{0} - 1s\,2p\,^3P_{2}$                  & $x$    & 3.18941 & 3.18941  & \\%1
Ca XVIII & $1s^2\,2s\,^2S_{1/2} - 1s\,2s\,2p\,(^3P)\,^2P_{3/2}$ & $s$  & 3.1902  & 3.1869   & 3.00(13) & 1.74(13) & 2.61(13) \\%23
Ca XVIII & $1s^2\,2s\,^2S_{1/2} - 1s\,2s\,2p\,(^3P)\,^2P_{1/2}$ & $t$  & 3.1912  & 3.19113  & 5.34(13) & 5.26(13) & 5.31(13)  \\%22
Ca XIX   & $1s^2\,^1S_{0} - 1s\,2p\,^3P_{1}$                  & $y$    & 3.19291 & 3.19291  & \\%2
Ar XVII  & $1s^2\,^1S_{0} - 1s\,4p\,^1P_{1}$                  & $w4$   & 3.1997  & 3.1997   &          \\%975
Ca XVIII & $1s^2\,2s\,^2S_{1/2} - 1s\,2s\,2p\,(^2P)\,^2P_{3/2}$ & $q$  & 3.20048 & 3.20048  & 1.06(13) & 2.68(12) & 8.51(11) \\%21
Ca XVIII & $1s^2\,2s\,^2S_{1/2} - 1s\,2s\,2p\,(^2P)\,^2P_{1/2}$ & $r$  & 3.20332 &  3.20332 & 3.27(13) & 2.88(13) & 3.18(13) \\%20
Ca XVIII & $1s^2\,2p\,^2P_{3/2} - 1s\,2p^2\,^2P_{3/2}$        & $a$    & 3.20332 & 3.20332  & 5.93(13) & 7.48(13) & 5.33(13) \\%14
Ar XVI   & $1s^2\,2p\,^2P - 1s 2p 5p\,(^3P)\,^2D$             & A5     & 3.2058  & 3.202    & 2.37(13) & 1.93(13) \\%2116,2117
Ca XVIII & $1s^2\,2p\,^2P_{1/2} - 1s\,2p^2\,^2D_{3/2}$        & $k$    & 3.20663 & 3.20663  & 2.34(14) & 2.48(14) & 2.27(14) \\%12
Ca XVIII & $1s^2\,2p\,^2P_{3/2} - 1s\,2p^2\,^2D_{5/2}$        & $j$    & 3.21095 & 3.21095  & 3.21(14) & 3.33(14) & 3.10(14) \\%10
Ca XIX  & $1s^2\,^1S_{0} - 1s\,2s\,^3S_{1}$                   & $z$    & 3.21095 & 3.21095    \\%3
Ca XVII  & $1s^2\,2s\,2p\,^3P_2 - 1s\,2s\,2p^2\,(^1S)\,^3S_1$ & G      & 3.2171  & 3.21724  & 5.59(13) & 4.43(13)   \\%564
Ca XVII  & $1s^2\,2s^2\,^1S_0 - 1s\,2s^2\,2p\,(^2S)\,^1P_1$   &$\beta$ & 3.221   & 3.2217   & 6.14(12) & 1.32(14)   \\%531+974
Ca XVIII & $1s^2\,2s\,^2S_{1/2} - 1s\,2s\,2p\,^4P_{3/2}$      & $u$    & 3.2264  & 3.2266   & 4.29(11) & 5.68(11) & 4.37(11) \\%19+2044
Ca XVII  & $1s^2\,2s\,2p\,^3P_0 - 1s\,2s\,2p^2\,(^1D)\,^3D_1$ & J      & 3.22645 & 3.22645  & 9.53(13) & 1.76(14)   \\%558
Ca XVII  & $1s^2\,2s\,2p\,^3P_1 - 1s\,2s\,2p^2\,(^1D)\,^3D_1$ & I?     & 3.2265  &          & 1.13(14) & 1.18(14)   \\%559
Ca XVII  & $1s^2\,2s\,2p\,^3P_2 - 1s\,2s\,2p^2\,(^3P)\,^3P_2$ & L      &3.22645? & 3.22645  & 1.30(14) & 2.35(14)    \\%557
Ca XVIII & $1s^2\,2s\,^2S_{1/2} - 1s\,2s\,2p\,^4P_{1/2}$      & $v$    & 3.2279  & 3.2277   & 6.56(10) & 1.19(12) & 3.3(10)  \\%18+2045
Ca XVII  & $1s^2\,2s\,2p\,^3P_1 - 1s\,2s\,2p^2\,(^1D)\,^3D_2$ & K      & 3.22827?& 3.22827  & 2.80(14) & 3.14(14)    \\%556
Ca XVII  & $1s^2\,2s\,2p\,^3P_2 - 1s\,2s\,2p^2\,(^1D)\,^3D_1$ &        & 3.2293  &          & 5.26(13) &          &             \\%553
Ca XVII  & $1s^2\,2s\,2p\,^3P_2 - 1s\,2s\,2p^2\,(^1D)\,^3D_3$ & N      & 3.23100?& 3.23100  & 3.65(14) & 3.99(14)    \\%552
Ca XVII  & $1s^2\,2s\,2p\,^1P_1 - 1s\,2s\,2p^2\,(^3P)\,^3P_2$ &        & 3.2331  &          & 5.28(13) &            \\%549
Ca XVII  & $1s^2\,2s\,2p\,^1P_1 - 1s\,2s\,2p^2\,(^1D)\,^1D_2$ & R      & 3.23677 & 3.23677  & 1.95(14) & 2.14(14)   \\%547
Ca XVI   & $1s^2\,2s^2\,2p\,^2P_{1/2} - 1s\,2s^2\,2p^2\,(^1D)\,^2D_{3/2}$&&3.2497&          & 2.94(14) &            \\%622
Ca XVI   & $1s^2\,2s^2\,2p\,^2P_{3/2} - 1s\,2s^2\,2p^2\,(^1D)\,^2D_{5/2}$&&3.2527&          & 3.78(14) &            \\%620
Ca XVIII & $1s^2\,2p\,^2P_{1/2} - 1s\,2s^2\,^2S_{1/2}$        & $p$    & 3.2636  &          & 5.79(12)$^b$& 5.38(12)&5.79(12) \\%25+2039
Ca XVIII & $1s^2\,2p\,^2P_{3/2} - 1s\,2s^2\,^2S_{1/2}$        & $o$    & 3.2688  &          & 8.27(12)$^b$& 7.62(12)&8.27(12)  \\%24+2038
Ar XVI   & $1s^2\,2p\,^2P  - 1s 2p 4p\,(^3P)\,^2D$            & A4     & 3.2731  & 3.270    & 4.47(13) & 3.12(13)   \\%2080,2081
Ca XV    & $1s^2\,2s^2\,2p^2\,^1D_2 - 1s\,2s^2\,2p^3\,(^2D)\,^1D_2$ &  & 3.2734  &          & 6.95(14)              \\%804

\\
\enddata
\tablenotetext{a}{Notation as follows: Ca XIX, Ca XVIII lines, \cite{gab72}; Ca XVII lines, \cite{see89}; Ar XVII lines, see text; Ar XVI lines, \cite{rice17}.}
\tablenotetext{b}{Adopted wavelengths: measured (col. 5) or Cowan $\lambda + 0.0032$~\AA. For Ca XVIII sats. $m$, $p$, $o$: \cite{bel82b}. For Ar XVIII Ly$\beta$ $\lambda$: \cite{eri77}. Intensity factor ($F_{\rm s}$, Eq. 4) for $p$, $o$: \cite{bel82b}.}
\tablenotetext{c}{Measured wavelengths: SOLFLEX \citep{see89} or Alcator C-Mod tokamak \citep{rice17}. }
\tablenotetext{d}{Intensity factor: Calculated here with Cowan code.}
\tablenotetext{e}{Intensity factor: \cite{vai78} or \cite{saf79}. Ar XVI lines: \cite{rice17}. }
\tablenotetext{f}{Intensity factor: \cite{bel82b}.}
\end{deluxetable}
%TTTTTTTTTTTTTTTTTTTTTTTTTTTTTTTTTTTTTTTTTTTTTTTTTTTTTTTTTTTTTTTTTTTTTTTTTTTTTTTTTTTTTTTTTTTTTTTTTTTTTTTTTTTTTTTTTTTTTTTTTTTTTTTTTTTTTTTTTTTT

% Subsection 3.4
\subsection{Ar XVII and Ar XVIII lines and satellites}
\label{sec:Ar_lines}

The \ion{Ar}{18} Ly$\beta$ and the \ion{Ar}{17} $w4$, $w5$, $w6$, $w7$, and (very weakly) $w8$ lines are evident in the higher-temperature DIOGENESS spectra and are included in Table~\ref{tab:line_wavelengths}. We used {\sc chianti} v.~8 for the intensities of the $w4$ and $w5$ lines and \cite{rice17} from Alcator C-Mod tokamak spectra measurements and theory for the line wavelengths. \ion{Ar}{16} dielectronic satellites associated with the \ion{Ar}{17} $w4$~--~$w8$ lines, although relatively weak,  are relevant for analysis of DIOGENESS spectra as some blend with the \ion{Ca}{19} line group. The \ion{Ar}{16} satellites have transitions $1s^2 nl - 1s nl n'l'$ where $nl$ are quantum numbers for the spectator electron and with $n'$ (the principal quantum number of the jumping electron) equal to 4 to 8. They have been observed and analyzed in Alcator C-Mod tokamak plasma spectra \citep{rice14,rice15}, with calculations of wavelengths and intensity factors $F_s$ given by \cite{rice17}. Generally, two of the most intense \ion{Ar}{16} satellite transitions, with transitions $1s^2\,2p\,^2P_{1/2} - 1s\, 2p \,n'p\,^2D_{3/2}$ and $1s^2\,2p\,^2P_{3/2} - 1s\, 2p\, n'p\,^2D_{5/2}$, dominate, making up what is essentially a single line feature (called by \cite{rice17} A4, A5, A6, A7, A8 for $n'=4$, 5, 6, 7, 8: this notation is used in Table~\ref{tab:line_wavelengths}). Here, wavelengths and $F_s$ values were calculated for these and other satellites with $n'$ up to 12 using the Cowan code, with 3.2~m\AA\ added to the Cowan wavelengths as with the calcium satellites. As can be seen from Table~\ref{tab:line_wavelengths}, there is good agreement for both the wavelengths and $F_s$ factors between the Cowan values and those of \cite{rice17} for the A4 to A8 satellites. Additional \ion{Ar}{16} satellite data were calculated with the Cowan code for satellites having transitions $1s^2 2l - 1s\,2l\,nl$ with $nl$ up to $12p$. Wavelengths and $F_s$ values of the \ion{Ar}{17} dielectronic satellites (transitions $1s 2s - 2s3p$, $1s 2p - 2p 3p$) associated with the \ion{Ar}{18} Ly$\beta$ were similarly calculated with the Cowan code and included in the synthetic spectra. Even the most intense of these satellites were found to make a very small contribution to the total spectrum.

Among the blends of argon lines or dielectronic satellites with the calcium lines, the most significant is that of the \ion{Ar}{17} $w4$ line (3.200~\AA) which is blended with \ion{Ca}{18} satellite $q$ (as was noted in {\em P78-1}\/ spectra: \cite{dos81b}); for the assumed Ar/Ca abundance ratio of 0.33, the lines are comparable in intensity over a wide temperature range. The \ion{Ar}{16} A5 satellite line feature is blended with \ion{Ca}{18} satellite $k$, although it makes a less than $\sim 2$\% contribution to line $k$. Both \ion{Ca}{18} satellites $q$ and $k$ have been extensively used as diagnostics of electron and ion temperature in {\em SMM}\/ BCS spectral analyses, so the blend of \ion{Ca}{18} satellite $q$ and the \ion{Ar}{17} $w4$ line should be taken into account. The \ion{Ar}{16} A6 feature is on the short-wavelength wing of the \ion{Ca}{19} $w$ line; the presence of a line at this wavelength was noticed in some {\em SMM}\/ BCS spectra but was not then previously identified.

% Subsection 3.5
\subsection{Continuum}

The continuum, made up of free--free, free--bound, and two-photon radiation, was calculated from {\sc chianti} routines and added to the line spectra, although in the analysis of the DIOGENESS spectra we compared only the line spectra as the DIOGENESS spectra have a continuous background which is mostly instrumental in origin. The continuum calculation used element abundances of \cite{fel92b} and ion fractions from the {\sc chianti} ionization equilibrium. Typically, free-bound emission is comparable to free--free emission over the 3.00~--~3.35~\AA\ range, but two-photon emission is over two orders of magnitude lower.

% Subsection 3.6
\subsection{Synthetic Spectra Displayed}

Figure~\ref{fig:theoretical_sp} shows a color-coded representation of synthetic 3.00--3.30~\AA\ spectra for temperatures in the range $4 - 30$~MK, with chief lines identified in the spectrum at the top, calculated for 13.8~MK. In these spectra, the line profile was taken to be Gaussian with width equal to the thermal Doppler broadening (FWHM $= 2.15 \times 10^{-7} \,\,T_e^{1/2}$~\AA\                                                                 for for calcium spectra in this wavelength range) convolved with the DIOGENESS rocking curve and a turbulent velocity of 100~km~s$^{-1}$ to take account of non-thermal broadening. For the lower temperatures (nearest the bottom of the figure), the \ion{Ca}{16} and \ion{Ca}{15} satellites at longer wavelengths predominate, but rapidly fade with increasing temperature. Correspondingly, the \ion{Ca}{19} line intensities increase with temperature until $\sim 16$~MK when the ion fraction of He-like Ca maximizes. The principal \ion{Ca}{18} satellites $q$, $r+a$, $k$ and $d13 + d15$ are also prominent at lower temperatures; for the dielectronically formed lines, their intensities decrease with an approximately $T_e^{-1}$ dependence relative to \ion{Ca}{19} line $w$. There is a similar decrease in the intensities of the \ion{Ca}{17} satellites on the long-wavelength side of \ion{Ca}{19} line $z$, reflecting the decreasing proportion of the associated ion fractions with temperature.

%***************************************************************************************************************************************************
% Section 4
\section{DIOGENESS AND THEORETICAL SPECTRA COMPARED}\label{sec:obs_calc_sp}

% Section 4.1
\subsection{DIOGENESS Solar Flare Spectra} \label{sec:DIOG}

\begin{figure}
\epsscale{1.0}
\plotone{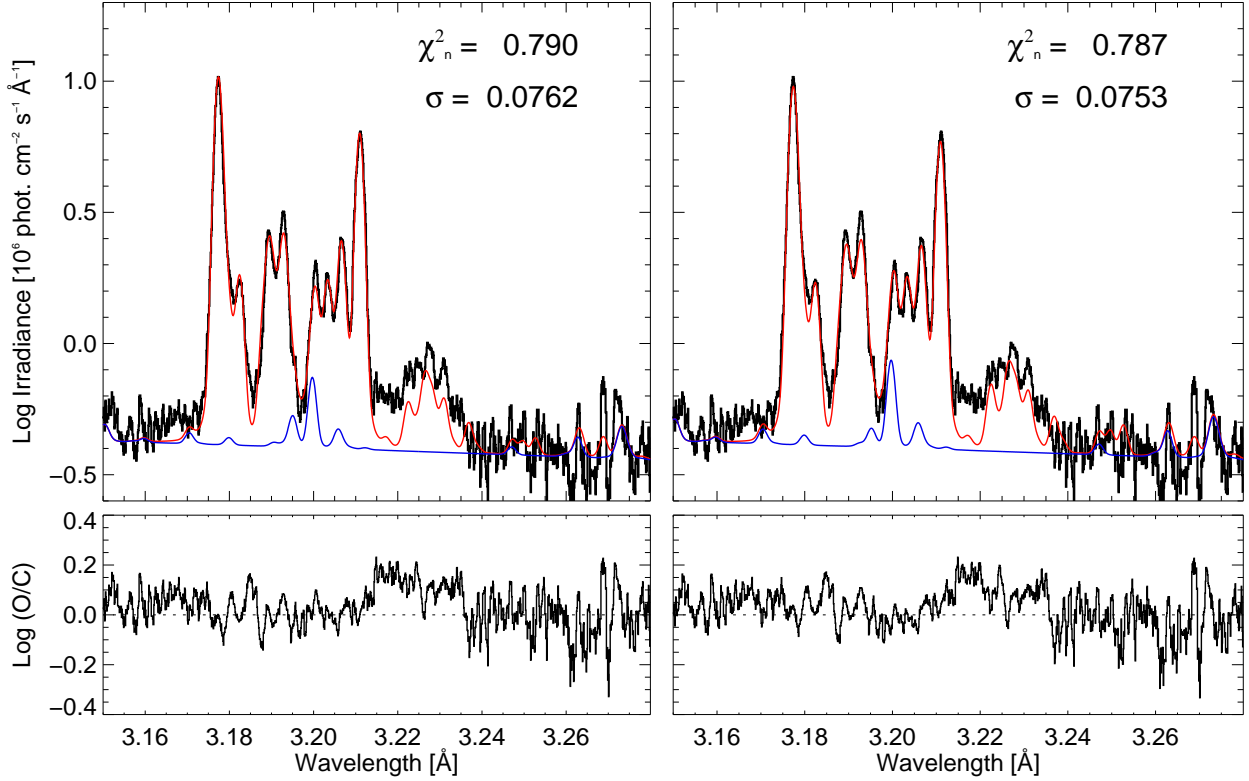}
%\vspace{25mm}
\caption{(Left) DIOGENESS spectrum for $T_{\rm GOES} = 12.8$~MK with best-fit theoretical spectrum (red continuous line) on a logarithmic scale, with contribution made by the argon spectrum alone (blue line); fit residuals are shown in the panel beneath. Line identifications may be found in Table~\ref{tab:line_wavelengths}. The ion fractions given in {\sc chianti} were used for the theoretical spectrum, and an Ar/Ca abundance ratio equal to 0.33 (Section~\ref{sec:Calc}). For this fit the reduced $\chi^2 = 0.79$; the largest residuals occur in the region of the \ion{Ca}{17} lines (3.21~--~3.23~\AA). (Right) As for the left panel but with the Ca$^{+16}$ and Ca$^{+17}$ ion fractions multiplied by 1.3 and 2.0 respectively. The \ion{Ca}{17} lines in the theoretical spectrum agree better, including line $\beta$ (3.2217\AA), as does the weak \ion{Ca}{16} line emission (3.24~--~3.25\AA), although $\chi^2$ is nearly identical. \label{fig:DIOG_sp_12pt75MK}}
\end{figure}

We compared the DIOGENESS spectra in Figure~\ref{fig:DIOG_av_sp_five_temps} with synthetic spectra calculated for temperatures equal to the appropriate value of $T_{\rm GOES}$. In the spectrum with the lowest temperature ($T_{\rm GOES} = 12.8$~MK), the low-temperature \ion{Ca}{17} satellite lines are most evident while the higher-temperature \ion{Ar}{17} and \ion{Ar}{18} are weak or non-existent. Fitting with theoretical spectra for this temperature should therefore best illustrate the reliability of the Cowan calculations of the calcium satellites. Figure~\ref{fig:DIOG_sp_12pt75MK} (left panel) shows the DIOGENESS spectrum with the best-fit theoretical spectrum; the background ``pedestal'' in the observed spectrum is fitted with an arbitrary background in the theoretical spectrum, and only the theoretical line spectrum without the continuum has been fitted to those in the observed spectrum. It is clear that the \ion{Ca}{17} lines in the 3.21~--~3.24~\AA, particularly \ion{Ca}{17} line $\beta$ (3.2217\AA), have intensities that are underestimated in the theoretical spectrum, and the weak emission in the 3.24~--~3.25\AA\ in the observed spectrum is not matched by the \ion{Ca}{16} line emission in the theoretical spectrum. The reduced $\chi^2$ for this fit is 0.79, and the residuals in the form log~$(O/C)$ ($O =$ observed, $C = $ calculated), which are plotted beneath the spectrum, are larger for the wavelength range of the \ion{Ca}{17} lines.

The \ion{Ca}{17} line $\beta$ is partly formed by inner-shell excitation from the Ca$^{+16}$ ion, the remaining \ion{Ca}{17} line emission by dielectronic recombination of Ca$^{+17}$ ions. Similarly, the weak \ion{Ca}{16} line emission is formed by dielectronic recombination of Ca$^{+16}$ ions. At a temperature of 12.8~MK, according to the {\sc chianti} ionization equilibrium calculations used here, the ion fractions of Ca$^{+17}$ and Ca$^{+16}$ are 0.13 and 0.02 respectively, decreasing sharply with temperature. In a separate work in preparation, we examine the likelihood of considerable errors in the ion fractions when small as a result of uncertainties in the ionization and recombination rates. These are generally based on theory with very few benchmark laboratory measurements. \cite{bry06} in an extensive discussion of ionization equilibria finds that the agreement of calculated and measured dielectronic recombination rates is 35\% or better, which is probably the typical agreement for other ionization or recombination rates. Our work indicates that an uncertainty of as small as 10\% in these rates leads to uncertainties in some ion fractions that are considerable, especially (as in the case of Ca$^{+17}$ and Ca$^{+16}$ at $T_e \gtrsim 12.8$~MK when the ion fractions are very small. In view of this, it seemed to us quite possible that the Ca$^{+16}$ and Ca$^{+17}$ ion fractions should be increased. The factors we used, equal to 1.3 and 2.0 respectively, were assumed to apply for the whole temperature range of the DIOGENESS spectra considered here (12.8~--~20.7~MK). This arbitrary assumption is not exact since it ignores the fact that the Ca$^{+18}$ ion fractions should be correspondingly adjusted, although since these fractions are nearly unity (between 0.82 and 0.72), the amount is slight. It is found that, with these adjusted ion fractions, the agreement between the DIOGENESS and theoretical spectra is much improved, as is shown by the fits to the \ion{Ca}{17} lines in Figure~\ref{fig:DIOG_sp_12pt75MK} (right panel). In addition, the weak emission at 3.24~--~3.25\AA\ in the observed spectrum can now be identified with the \ion{Ca}{16} dielectronic lines. (Table~\ref{tab:line_wavelengths} gives the two most intense lines.) We found, however, that the reduced $\chi^2$ in this fit (0.79) is almost identical to the fit in Figure~\ref{fig:DIOG_sp_12pt75MK} (left panel).

%FFFFFFFFFFFFFFFFFFFFFFFFFFFFFFFFFFFFFFFFFFFFFFFFFFFFFFFFFFFFFFFFFFFFFFFFFFFFFFFFFFFFFFFFFFFFFFFFFFFFFFFFFFFFFFFFFFFFFFFFFFFFFFFFFFFFFFFFFFFF
% Figure 5: Two DIOG spectra for T=12.75MK with fits - without adjusted ion fractions & with.
%FFFFFFFFFFFFFFFFFFFFFFFFFFFFFFFFFFFFFFFFFFFFFFFFFFFFFFFFFFFFFFFFFFFFFFFFFFFFFFFFFFFFFFFFFFFFFFFFFFFFFFFFFFFFFFFFFFFFFFFFFFFFFFFFFFFFFFFFFFFF

It is clear from Figure~\ref{fig:DIOG_sp_12pt75MK} that the \ion{Ar}{17} lines make a substantial contribution (blue line in figure). At 12.8~MK, only the \ion{Ar}{17} $w5$ is prominent on the short-wavelength side of the \ion{Ca}{19} line $w$, where there is agreement of the theoretical spectrum with the observed. The $w4$ line is not evident because of the blend with the \ion{Ca}{18} satellite $q$ (3.2005~\AA), but its intensity calculated on the assumption of an Ar/Ca abundance of 0.33 is 25\% of the total intensity of this blended feature. At wavelengths longer than \ion{Ca}{19} line $z$ (3.211~\AA), the \ion{Ar}{16} satellites seen in Alcator C-Mod tokamak spectra, including the A4 line feature, are visible, as are the \ion{Ca}{18} $o$ and $p$ satellites. This is the first time these lines as well as the \ion{Ca}{16} satellites have been noted in astrophysical spectra.

%FFFFFFFFFFFFFFFFFFFFFFFFFFFFFFFFFFFFFFFFFFFFFFFFFFFFFFFFFFFFFFFFFFFFFFFFFFFFFFFFFFFFFFFFFFFFFFFFFFFFFFFFFFFFFFFFFFFFFFFFFFFFFFFFFFFFFFFFFFFF
% Figure 6: Four DIOG spectra with fits, wide range
\begin{figure}
\epsscale{1.}
\plotone{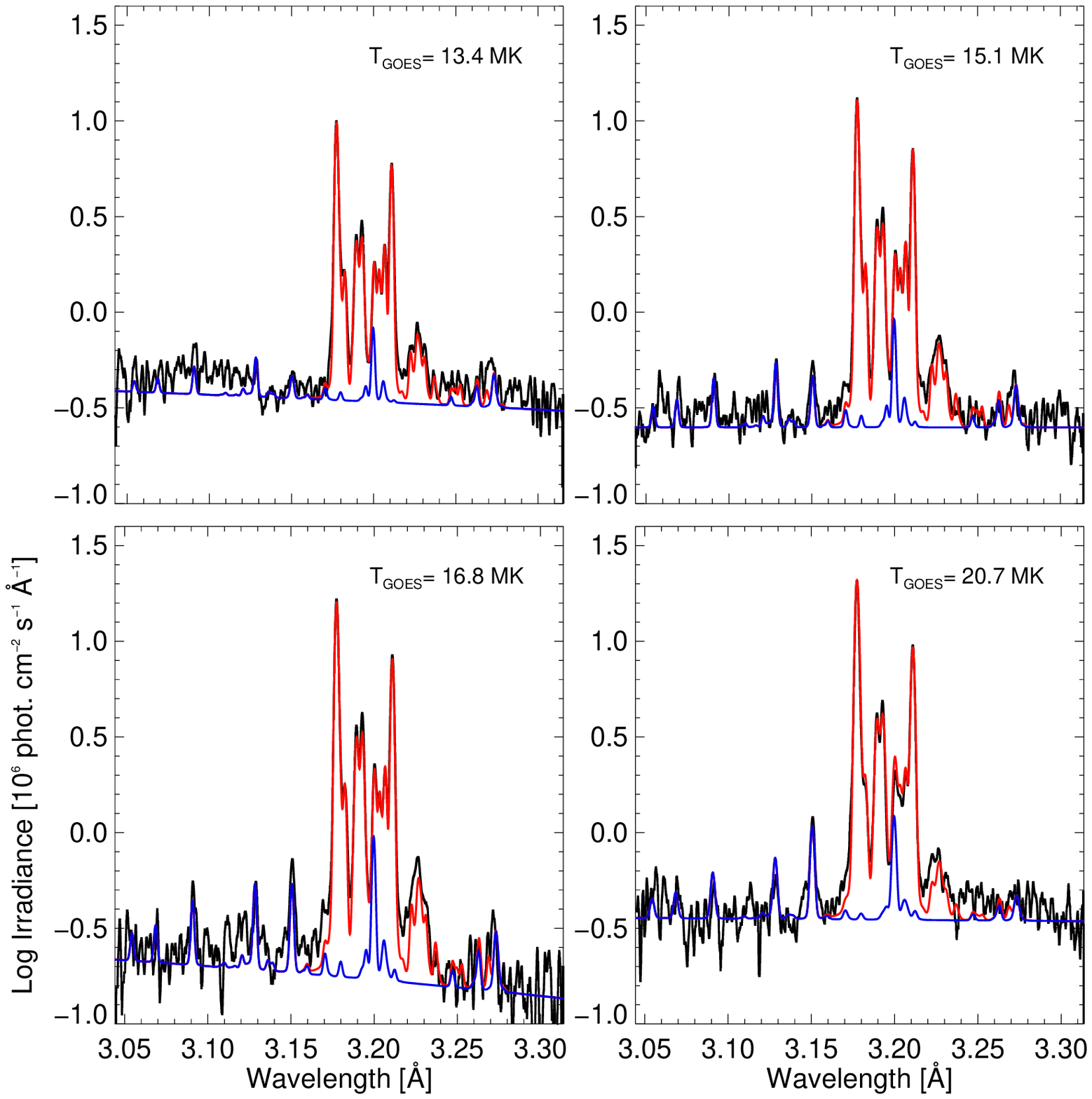}
\caption{DIOGENESS spectra in the range 3.05~--~3.30~\AA\ for $T_{\rm GOES} = 13.4$~MK, 14.8~MK, 16.8~MK, and 20.7~MK with best-fit theoretical spectra (red continuous line) with argon spectrum (blue line). The Ca$^{+16}$ and Ca$^{+17}$ ion fractions were multiplied by 1.3 and 2.0 respectively as in Figure~\ref{fig:DIOG_sp_12pt75MK} (right panel). The values of reduced $\chi^2$ for these fits are 1.06, 1.13, 1.97, and 3.30.  \label{fig:fits_four_av_sp_wide}}
\end{figure}
%FFFFFFFFFFFFFFFFFFFFFFFFFFFFFFFFFFFFFFFFFFFFFFFFFFFFFFFFFFFFFFFFFFFFFFFFFFFFFFFFFFFFFFFFFFFFFFFFFFFFFFFFFFFFFFFFFFFFFFFFFFFFFFFFFFFFFFFFFFFF

%FFFFFFFFFFFFFFFFFFFFFFFFFFFFFFFFFFFFFFFFFFFFFFFFFFFFFFFFFFFFFFFFFFFFFFFFFFFFFFFFFFFFFFFFFFFFFFFFFFFFFFFFFFFFFFFFFFFFFFFFFFFFFFFFFFFFFFFFFFFF
% Figure 7: Four DIOG spectra with fits, narrow range
\begin{figure}
\epsscale{1.}
\plotone{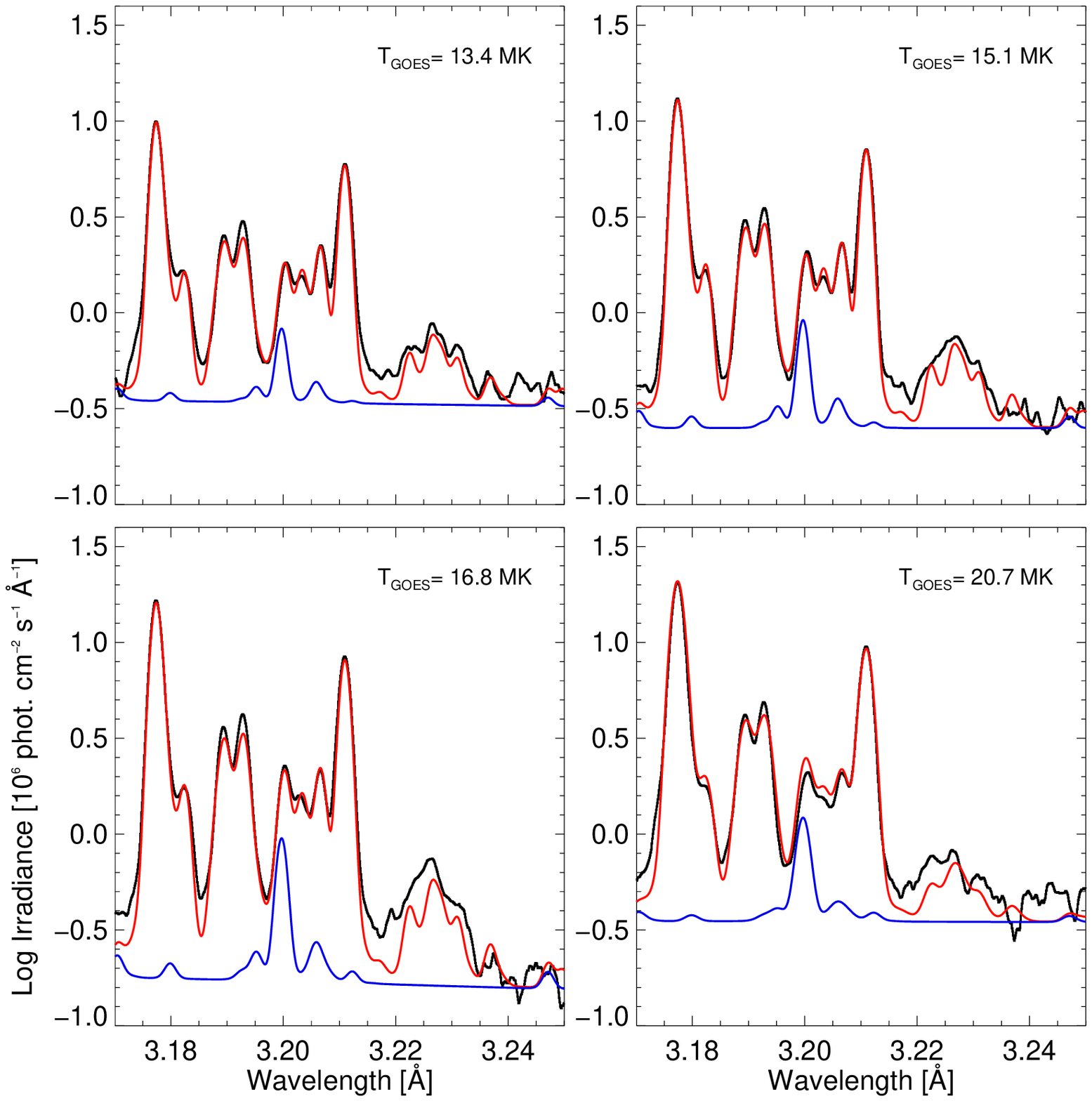}
\caption{DIOGENESS spectra (3.17~--~3.25~\AA) for $T_{\rm GOES} = 13.4$~MK, 14.8~MK, 16.8~MK, and 20.7~MK with best-fit theoretical spectra (red continuous line) with argon spectrum (blue line) for a wavelength range that includes only the region around the Ca XIX lines and Ca XVIII and Ca XVII satellites. The Ca$^{+16}$ and Ca$^{+17}$ ion fractions were multiplied by 1.3 and 2.0 respectively as in Figure~\ref{fig:DIOG_sp_12pt75MK}. The values of reduced $\chi^2$ for these fits are 2.41, 3.23, 5.84, and 9.86. \label{fig:fits_four_av_sp_narrow}}
\end{figure}
%FFFFFFFFFFFFFFFFFFFFFFFFFFFFFFFFFFFFFFFFFFFFFFFFFFFFFFFFFFFFFFFFFFFFFFFFFFFFFFFFFFFFFFFFFFFFFFFFFFFFFFFFFFFFFFFFFFFFFFFFFFFFFFFFFFFFFFFFFFFF

With the adjustments in \ion{Ca}{17} and \ion{Ca}{16} ion fractions from the fits at a temperature of 12.8~MK, we fitted the spectra for the remaining temperatures (13.4~MK, 14.8~MK, 16.8~MK, 20.7~MK), obtaining satisfactory fits for the first two with values of reduced $\chi^2$ of 1.06 and 1.13. Figure~\ref{fig:fits_four_av_sp_wide} shows these four spectra, which have the same color code as Figure~\ref{fig:DIOG_sp_12pt75MK}.  For the two higher temperatures, reduced $\chi^2$ was 1.97 and 3.33, the slightly worse fit apparently arising from the noisier background which may be due to enhanced fluorescence. At the higher temperatures, the argon lines are more prominent, particularly the \ion{Ar}{18} Ly-$\beta$ line at 3.151~\AA, which are well fitted so verifying the {\sc chianti} atomic data on which the line intensities are based. The measured wavelengths of \cite{rice14} agree with the DIOGENESS wavelengths, which therefore validates the DIOGENESS wavelength scale.

When narrower (3.17~--~3.25~\AA) wavelength ranges are chosen, the slightly discrepant intensities of the \ion{Ca}{19} line $y$ and to a smaller extent line $x$, in which the observed intensities are greater than the theoretical, are more apparent (Figure~\ref{fig:fits_four_av_sp_narrow}). This has been recognized in the past (e.g. Figure~5 of \cite{bel82b}) and also in fits to the equivalent \ion{Fe}{25} lines \citep{lem84}. High-$n$ \ion{Ca}{18} satellites which converge on to these lines could possibly account for this, but in the synthetic spectral program this has been taken account of using the procedure of \cite{bel79a} in which the total intensity of each satellite group is proportional to $n^{-3}$ and the peak wavelength is displaced from either $x$ or $y$ by an amount $\Delta \lambda$ which is proportional to $n^{-3}$. We used the Cowan calculations for the $n=6$ satellites, calculated specifically, for the intensity and $\Delta \lambda$ for satellites with $n=7$ to $n=16$, as in \cite{bel79a}. However, the amount of emission added was extremely small and did not account for the observed discrepancy. Unless it can be attributed to the \cite{bel82b} calculated collision strengths rather than more accurate close coupling data (see Section~\ref{sec:CaXIX}), the reason for the discrepancy remains unknown. In Figure~\ref{fig:fits_four_av_sp_narrow}, the \ion{Ca}{17} line group is displayed to better advantage. Most of the satellite structure is accounted for by the present calculations, which give similar results to those of \cite{see89}. At higher temperatures, \ion{Ca}{18} satellite $u$ becomes more prominent relative to the \ion{Ca}{17} satellites, as is indicated in the DIOGENESS spectrum for 20.7~MK.

% Section 4.2
\subsection{Solar Flare Spectra from P78-1 SOLFLEX} \label{sec:P78}

\begin{figure}
\epsscale{.70}
\plotone{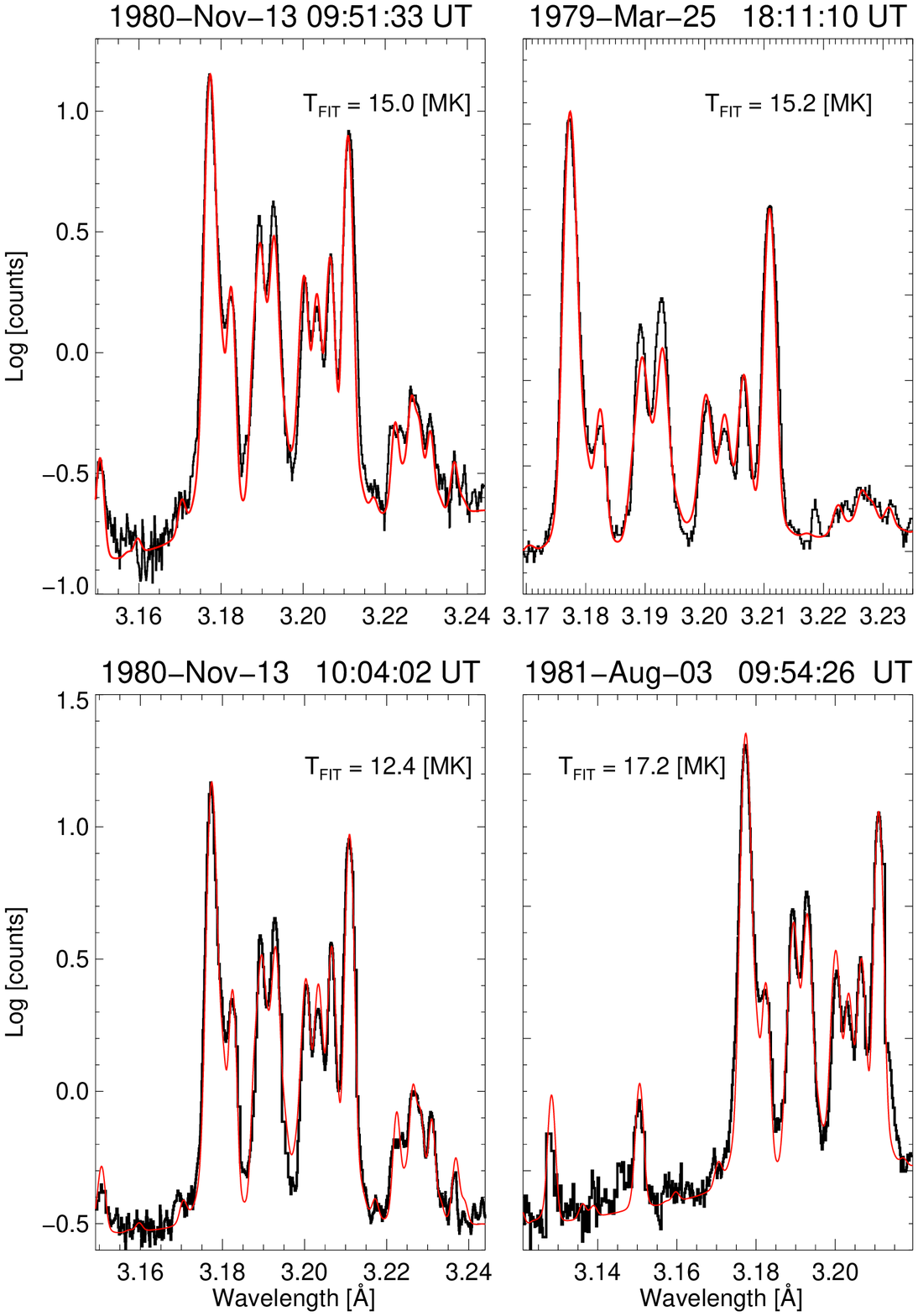}
%\vspace{10mm}
\caption{Four spectra from the {\em P78-1} SOLFLEX instrument during the flares of 1979 March 25, 1980 August 3 and 1981 November~13 (black lines) with spectral fits (red), with temperatures (indicated) from the \ion{Ca}{18} $k$ line to \ion{Ca}{19} $w$ line ratio. The spectra were digitized from figures in \cite{dos79} and \cite{see89}. The theoretical spectra (in red) are based on the $k/w$ temperature and Ar/Ca abundance ratio of 0.33, with adjustments to the Ca$^{+17}$ and Ca$^{+16}$ ion fractions as discussed in the text. \label{fig:P78_sp_fits} }
\end{figure}

The computer program generating synthetic spectra in the region of the \ion{Ca}{19} X-ray lines was written primarily to match the DIOGENESS spectra obtained in the large flare of 2001 August~25, but other spectra in which the \ion{Ca}{19} lines and lower-ionization satellites are visible can be examined with the same program. These include high-resolution SOLFLEX spectra of the calcium lines from the {\em P78-1} spacecraft discussed by \cite{dos79} and \cite{see89}. Four of these spectra were digitised and analyzed with the synthetic spectrum program including the same adjustments to the Ca$^{+16}$ and Ca$^{+17}$ ion fractions as discussed in Section~\ref{sec:DIOG}. Figure~\ref{fig:P78_sp_fits} shows examples of four spectra from three large flares, with temperatures estimated from the intensity ratio of \ion{Ca}{18} satellite $k$ to \ion{Ca}{19} $w$. The line widths were broadened by a turbulent velocity of 60~km~s$^{-1}$ in all cases.  With these isothermal fits, almost all the principal line features are reproduced in the synthetic spectra, including the main lines within the \ion{Ca}{17} line group at 3.21~--~3.24~\AA, given in Table~\ref{tab:line_wavelengths}. Argon lines are apparent, particularly the \ion{Ar}{18} Ly-$\beta$ line at 3.151~\AA\ and the weak \ion{Ar}{16} A6 satellite feature (3.170~\AA) on the short-wavelength side of \ion{Ca}{19} line $w$. The intensities of these lines in the theoretical spectra are very similar to the observed, so the Ar/Ca abundance ratio of 0.33 appears to apply to these spectra also. However, the mismatch of the \ion{Ca}{19} lines $x$ and $y$, as in the DIOGENESS spectra, occurs here, so indicating a problem in the calculation of the excitation of these lines and {\em not}\/ the observations.

%FFFFFFFFFFFFFFFFFFFFFFFFFFFFFFFFFFFFFFFFFFFFFFFFFFFFFFFFFFFFFFFFFFFFFFFFFFFFFFFFFFFFFFFFFFFFFFFFFFFFFFFFFFFFFFFFFFFFFFFFFFFFFFFFFFFFFFFFFFFF
% Figure 8

% Section 4.3
\subsection{Alcator C-Mod Tokamak Spectra} \label{sec:Alcator}

\begin{figure}
\epsscale{.70}
\plotone{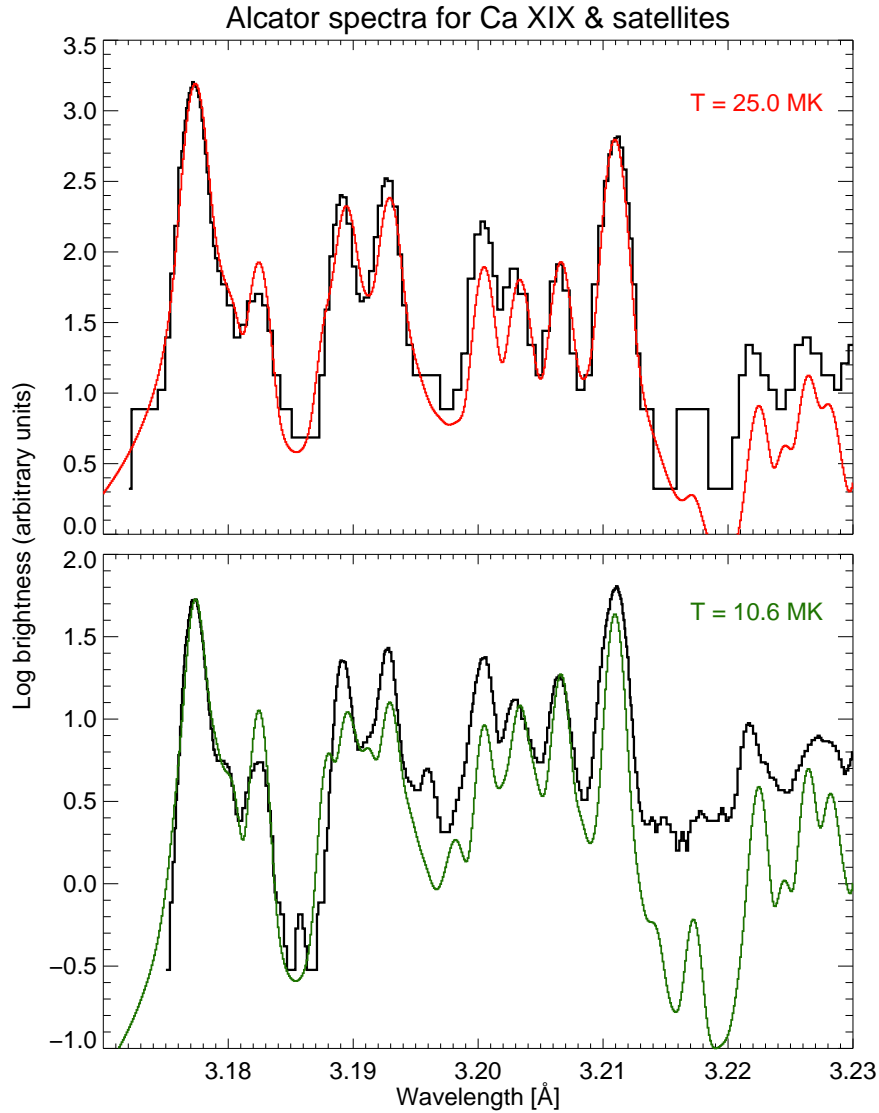}
%\vspace{10mm}
\caption{Two spectra in the neighborhood of the \ion{Ca}{19} lines taken from the Alcator~C-Mod tokamak, shown by \cite{rice14} and digitized here; the top panel is for the spectrometer viewing the plasma center, the lower panel at a distance half-way between the plasma center and its edge. The fitted synthetic spectra (red curves) have temperatures (in the figure legend) estimated from fitting the $k/w$ line ratios as in Figure~\ref{fig:P78_sp_fits}.  \label{fig:Alcator_fits} }
\end{figure}

High-resolution spectra of the calcium lines discussed here were obtained from an X-ray spectrometer viewing plasma produced in the Alcator C-Mod tokamak. They have been analyzed by \cite{rice14} and \cite{rice17}. The measured electron density, $\sim 10^{14}$~cm$^{-3}$, at the center of the plasma is about a factor of 100 higher than those commonly measured in solar flare plasmas but the ionization and excitation conditions are similar. Two spectra shown in figures by \cite{rice14} were digitized and compared with synthetic spectra using only the calcium line spectra without the argon lines; they are shown in Figure~\ref{fig:Alcator_fits}. The top panel shows the spectrum viewed by the spectrometer through a central chord; the central temperature of the plasma was measured to be 3.5~keV (41~MK), much higher than any of the DIOGENESS or {\em P78-1}\/ SOLFLEX solar flare spectra, though the presence of \ion{Ca}{18} satellites $q$, $r/a$, and $k$ and even \ion{Ca}{17} lines at wavelengths 3.22~--~3.23~\AA\ suggests that lower-temperature plasma is present also, presumably along the line of sight of the spectrometer, either side of the plasma center. The observed line widths appear to be smaller than the thermal Doppler broadened values, which may indicate that the ion temperature is smaller than the electron temperature. A synthetic spectrum with a temperature of 25~MK approximately matches the observed, but \ion{Ca}{17} satellite $q$ is observed to be more intense than calculated, as are the \ion{Ca}{17} satellites; this may be due to the assumption of a single temperature whereas lower-temperature plasma is also being viewed. In Figure~\ref{fig:Alcator_fits} (lower panel), the spectrum was obtained by viewing the plasma at a distance of 0.5 times the distance from the plasma center to its edge and therefore shows a much lower temperature. We found an approximate fit with the synthetic spectrum having a temperature of 10.7~MK. The mismatches (e.g. in the \ion{Ca}{18} satellite $q$ which is largely formed by inner-shell excitation rather than dielectronic recombination) may arise from lower-temperature plasma in the line of sight as in the higher-temperature spectrum.

%FFFFFFFFFFFFFFFFFFFFFFFFFFFFFFFFFFFFFFFFFFFFFFFFFFFFFFFFFFFFFFFFFFFFFFFFFFFFFFFFFFFFFFFFFFFFFFFFFFFFFFFFFFFFFFFFFFFFFFFFFFFFFFFFFFFFFFFFFFFF
% Figure 9

% Section 5
\section{RELATION OF \ion{Ca}{19} AND {\em GOES}\/ TEMPERATURES}\label{sec:GOES_Ts}

We have assumed in the above analysis that an isothermal plasma with temperature $T_{\rm GOES}$ can describe DIOGENESS spectra adequately. Our experience with RESIK spectra has shown that this is a good approximation for describing \ion{K}{19} and \ion{Ar}{17} line emission including satellites (seen in RESIK channels~1 and 2) during flare decays \citep{jsyl10a,jsyl10b}. An isothermal assumption was found to be less satisfactory during rapidly rising phases of flares. This agrees with an analysis of the late, nearly isothermal, stages of long-duration flares seen with the {\em Yohkoh}\/ Bragg Crystal Spectrometer \citep{phi05}. However, RESIK flare observations of the lower-temperature \ion{S}{15}, \ion{S}{16}, and \ion{Si}{14} line emission indicated a less satisfactory fit on an isothermal assumption \citep{bsyl13}, and instead a differential emission measure procedure was used to derive S and Si abundances \citep{bsyl15}; $T_{\rm GOES}$ was found to be near the higher temperature range of the differential emission measure. A dependence on atomic number is thus suggested, an isothermal assumption with temperature equal to $T_{\rm GOES}$ adequately describing helium-like Ar ($Z=18$) and  K ($Z=19$) line spectra during flare decays but not spectra from helium-like Si ($Z=14$), S ($Z=16$), and lower-$Z$ elements for which lower-temperature plasma contributes to the line emission.

\begin{figure}
\epsscale{1.0}
\plotone{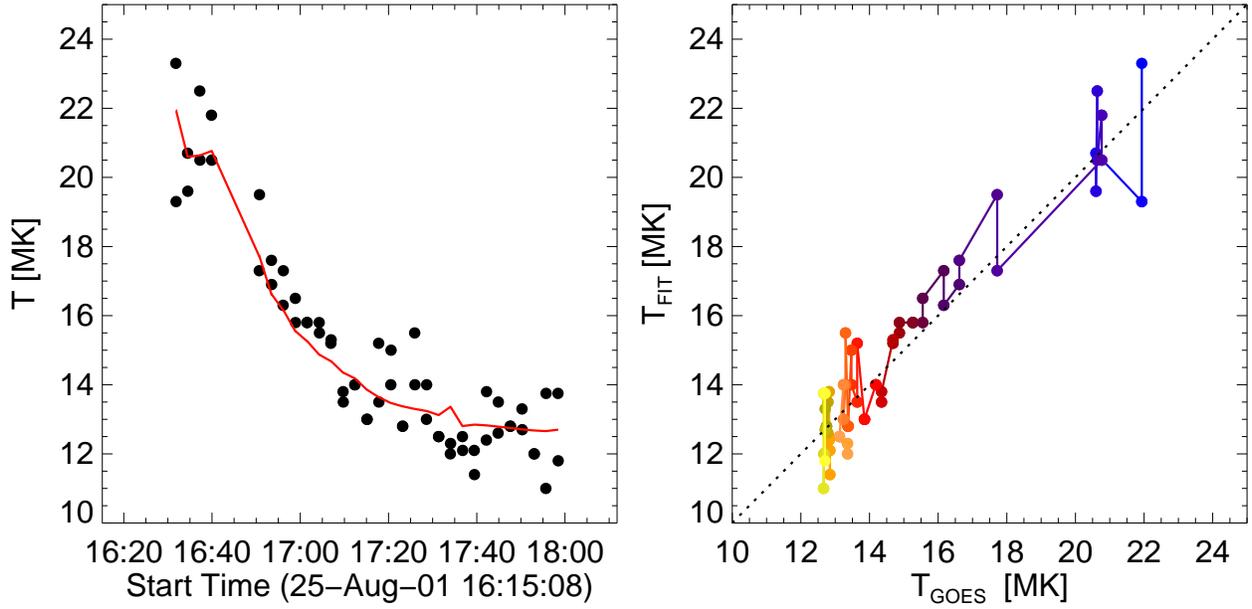}
%\vspace{10mm}
\caption{Left: Temperature ($T_{\rm FIT}$) (black dots) estimated from the fit to each DIOGENESS spectrum for spectra starting at scan 5 during the 2001 August~25 flare and $T_{\rm GOES}$ plotted against time. Right: $T_{\rm FIT}$ plotted against $T_{\rm GOES}$ (colored dots) for the same time period, blue dots indicating higher temperatures (earlier times), lighter colors lower temperatures (later times). The dashed line indicates equal temperatures.   \label{fig:TFIT_Tgoes}}
\end{figure}

Here we examine the validity of an isothermal assumption for DIOGENESS \ion{Ca}{19} spectra during the 2001 August~25 flare. We derived $T_e$ from the temperature of the best fit ($T_{\rm FIT}$) synthetic spectrum to each of sixty DIOGENESS spectra starting from scan 5 (see Figure~\ref{fig:DIOG_1-4_GOES_time}, near the flare maximum and the maximum of the {\em Yohkoh}\/ HXT signal). The values of $T_{\rm FIT}$ are plotted (black points) in Figure~\ref{fig:TFIT_Tgoes} (left panel) against time (UT), with corresponding values of $T_{\rm GOES}$ (red line). The precision of the later values of $T_{\rm FIT}$ is reduced as the flare spectra become less intense. A near-exact equality of the two temperatures is observed (Figure~\ref{fig:TFIT_Tgoes}, right panel), the color code indicating temperature as well as time, with the higher temperatures measured at the earlier times. This is despite the fact that the earliest times include some non-thermal emission, as shown by  HXT, which does not appear to affect the emission in either of the {\em GOES}\/ channels. The use of $T_{\rm GOES}$ therefore appears to be justified by this analysis, which is consistent with our earlier work on \ion{K}{19} and \ion{Ar}{17} line emission observed by RESIK.

%FFFFFFFFFFFFFFFFFFFFFFFFFFFFFFFFFFFFFFFFFFFFFFFFFFFFFFFFFFFFFFFFFFFFFFFFFFFFFFFFFFFFFFFFFFFFFFFFFFFFFFFFFFFFFFFFFFFFFFFFFFFFFFFFFFFFFFFFFFFF
% Figure 10: T(fit) and T(GOES) vs. time for many flare & July 5, 1980 flare. "Fig. 12"

% Section 5
\section{SUMMARY AND CONCLUSIONS}\label{sec:summary}

We have here described X-ray spectra in the range of channels~1 and 4 of the DIOGENESS spectrometer on {\em CORONAS-F}\/ during a powerful (X5) flare on 2001 August~25. The wavelength range, approximately 3.05~--~3.35~\AA, includes prominent lines of \ion{Ca}{19} and associated dielectronic satellites and ionized argon (\ion{Ar}{17}, \ion{Ar}{18}) lines including \ion{Ar}{16} dielectronic satellites. The spectra were divided into five groups according to the value of $T_{\rm GOES}$, the temperature obtained from the emission ratio of the two channels of {\em GOES}, which ranged from 12.8~MK to 20.7~MK. A computer program was used to synthesize the line spectrum based on atomic data of \cite{bel82b} for the \ion{Ca}{19} lines, the Cowan Hartree--Fock code for the wavelengths and intensity factors for dielectronic satellites of \ion{Ca}{18}, \ion{Ca}{17}, and \ion{Ca}{16}, atomic data for the \ion{Ar}{17} and \ion{Ar}{18} lines from {\sc chianti}, and the Cowan code for associated dielectronic satellites. Table~\ref{tab:line_wavelengths} lists the more intense lines included in the synthetic spectral calculation which included over 2000 lines. Comparison with the DIOGENESS spectra averaged over the five temperature intervals shows good agreement for the \ion{Ca}{19} and \ion{Ca}{18} satellites, but the \ion{Ca}{17} and weaker \ion{Ca}{16} satellites were found to be more intense in the DIOGENESS spectra. The argon lines were generally well fitted by the synthetic spectra. By increasing the Ca$^{+17}$ and Ca$^{+16}$ ion fractions by factors of 1.3 and 2.0 respectively, much improved agreement was achieved. This adjustment of the ion fractions is justified by the fact that the Ca$^{+17}$ and Ca$^{+16}$ ion fractions are small at all relevant temperatures, and a roughly 10\% uncertainty in ionization and recombination rates would easily explain such increases. Comparison of synthetic spectra with those from the {\em P78-1}\/ SOLFLEX instrument (Figure~\ref{fig:P78_sp_fits}) shows good agreement, although comparison with the calcium spectra produced in the Alcator C-Mod tokamak are slightly less satisfactory, the explanation probably being the multi-thermal and non-equilibrium or non-stationary nature of the plasma.

From the synthetic spectra, we find that the overlap of ionized calcium and argon lines in the spectral range of interest is important in at least one case, the blend of the \ion{Ca}{18} satellite $q$ (3.2005~AA), commonly used to diagnose the non-equilibrium state of flare plasmas \citep{bel82b}, with the \ion{Ar}{17} $w4$ line (3.1997~\AA), as was pointed out by \cite{dos81b}. With an assumed Ar/Ca ratio of 0.33 (based on measurements by \cite{jsyl98} and \cite{jsyl10b}), the contribution made by the \ion{Ar}{17} line is comparable to the $q$ line. As pointed out by \cite{jsyl98}, the calcium abundance varies from flare to flare, so the blend may be more or less serious according as the calcium abundance is less or more; an extreme case has been pointed out by \cite{dos16} who find from ultraviolet line emission near a sunspot that the Ar/Ca abundance ratio may be more than photospheric, equal to around 1.4. On the other hand, a large enhancement of the calcium abundance might occur in flares on active dwarf M stars, if the FIP effect is larger than that apparently operating in the solar atmosphere; thus, the Ar/Ca abundance ratio may be less than our assumed value, reducing the effect of argon line blends. It should be noted also that line blends occur in other X-ray spectral regions, e.g. with the \ion{S}{15} $w6$ line (3.998~\AA) very near the \ion{Ar}{17} lines at 3.969~--~3.994~\AA. These may be relevant to observations of stellar coronae taken with the {\em Chandra}\/ High Energy Transmission Grating Spectrometer.

Weaker \ion{Ar}{16} satellites in DIOGENESS spectra make small contributions, e.g. the A5 feature (3.2058~\AA) makes a few percent contribution to \ion{Ca}{18} line $k$ (3.2066~\AA), and the A6 feature (3.1704~\AA) may appear on the short-wavelength side of the \ion{Ca}{19} line $w$ (3.1774~AA). With the very high sensitivity and spectral resolution of DIOGENESS, the weak \ion{Ar}{16} satellite features A6 and A4 (3.273~\AA) and the \ion{Ca}{18} $p$ and $o$ satellites (3.2636~\AA, 3.2688~\AA) are evident, particularly in the spectrum with $T_{\rm GOES} = 16.8$~MK. These lines have not been recorded in solar flare spectra before.

We have also examined the assumption of isothermal plasma to describe the DIOGENESS spectra, plotting the temperature ($T_{\rm FIT}$) of the best-fit synthetic spectrum to each DIOGENESS spectrum from scan 5 to the end of the observations against $T_{\rm GOES}$, finding an almost exact equality. An isothermal description of \ion{Ca}{19} emission is therefore valid, with the temperature given by $T_{\rm GOES}$.

Although solar flare X-ray spectra in the region of the \ion{Ca}{19} lines have not recently been observed, analysis of the extensive archive of {\em SMM}\/ BCS spectra is now possible with application of the synthetic spectral code described here. In particular, it will be possible to re-analyze spectra used to derive the flare calcium abundance and deduce more exactly the relation of the variability of this abundance with flare type; this analysis is now proceeding. Future spacecraft instrumentation -- the ChemiX spectrometer, due for launch on {\em Interhelioprobe}\/ in 2025 or 2026 -- will benefit from the synthetic spectral procedure described here.

\acknowledgments

We acknowledge financial support from the Polish National Science Centre grant numbers UMO-2013-11/B/ST9/00234 and UMO-2017/25/B/ST9/01821.
We acknowledge the help of Jarek Baka{\l}a in digitizing the {\em P78-1}\/ and Alcator spectra, \.{Z}aneta Szaforz for calculating the total quartz crystal reflectivities using the XOP package, and Marek \'{S}t\k{e}\'{s}licki for reformating the DIOGENESS telemetry. The {\sc chianti} atomic database and code is a collaborative project involving George Mason University, University of Michigan (USA), and University of Cambridge (UK).

\vspace{5mm}
{\em Facilities:} \facility{GOES}, \facility{CORONAS/DIOGENESS}, \facility{CORONAS/RESIK}, \facility{Yohkoh/HXT}, \facility{Yohkoh/SXT}

%\software{IDL}

\bibliographystyle{apj}

\bibliography{RESIK}

\begin{thebibliography}{42}
\expandafter\ifx\csname natexlab\endcsname\relax\def\natexlab#1{#1}\fi

\bibitem[{{Aggarwal} \& {Keenan}(2012)}]{agg12}
{Aggarwal}, K.~M., \& {Keenan}, F.~P. 2012, \physscr, 85, 025306

\bibitem[{{Asplund} {et~al.}(2009){Asplund}, {Grevesse}, {Sauval}, \&
  {Scott}}]{asp09}
{Asplund}, M., {Grevesse}, N., {Sauval}, A.~J., \& {Scott}, P. 2009, Ann. Rev.
  Astr. Astroph., 47, 481

\bibitem[{{Bely-Dubau} {et~al.}(1982){Bely-Dubau}, {Faucher}, {Steenman-Clark},
  {Dubau}, {Loulergue}, {Gabriel}, {Antonucci}, {Volonte}, \&
  {Rapley}}]{bel82b}
{Bely-Dubau}, F., {Faucher}, P., {Steenman-Clark}, L., {Dubau}, J.,
  {Loulergue}, M., {Gabriel}, A.~H., {Antonucci}, E., {Volonte}, S., \&
  {Rapley}, C.~G. 1982, \mnras, 201, 1155

\bibitem[{{Bely-Dubau} {et~al.}(1979){Bely-Dubau}, {Gabriel}, \&
  {Volonte}}]{bel79a}
{Bely-Dubau}, F., {Gabriel}, A.~H., \& {Volonte}, S. 1979, \mnras, 189, 801

\bibitem[{{Bryans} {et~al.}(2006){Bryans}, {Badnell}, {Gorczyca}, {Laming},
  {Mitthumsiri}, \& {Savin}}]{bry06}
{Bryans}, P., {Badnell}, N.~R., {Gorczyca}, T.~W., {Laming}, J.~M.,
  {Mitthumsiri}, W., \& {Savin}, D.~W. 2006, \apjs, 167, 343

\bibitem[{{Bryans} {et~al.}(2009){Bryans}, {Landi}, \& {Savin}}]{bry09}
{Bryans}, P., {Landi}, E., \& {Savin}, D.~W. 2009, \apj, 691, 1540

\bibitem[{{Burgess} \& {Tully}(1992)}]{bur92}
{Burgess}, A., \& {Tully}, J.~A. 1992, \aap, 254, 436

\bibitem[{{Cowan}(1981)}]{cow81}
{Cowan}, R.~D. 1981, {The theory of atomic structure and spectra} (Berkeley:
  University of California Press, 1981)

\bibitem[{{Del Zanna} {et~al.}(2015){Del Zanna}, {Dere}, {Young}, {Landi}, \&
  {Mason}}]{delz15}
{Del Zanna}, G., {Dere}, K.~P., {Young}, P.~R., {Landi}, E., \& {Mason}, H.~E.
  2015, \aap, 582, A56

\bibitem[{{Dere} {et~al.}(1997){Dere}, {Landi}, {Mason}, {Monsignori Fossi}, \&
  {Young}}]{der97}
{Dere}, K.~P., {Landi}, E., {Mason}, H.~E., {Monsignori Fossi}, B.~C., \&
  {Young}, P.~R. 1997, \aaps, 125, 149

\bibitem[{{Doschek} \& {Feldman}(1981)}]{dos81b}
{Doschek}, G.~A., \& {Feldman}, U. 1981, \apj, 251, 792

\bibitem[{{Doschek} {et~al.}(1979){Doschek}, {Kreplin}, \& {Feldman}}]{dos79}
{Doschek}, G.~A., {Kreplin}, R.~W., \& {Feldman}, U. 1979, \apjl, 233, L157

\bibitem[{{Doschek} \& {Warren}(2016)}]{dos16}
{Doschek}, G.~A., \& {Warren}, H.~P. 2016, \apj, 825, 36

\bibitem[{{Erickson}(1977)}]{eri77}
{Erickson}, G.~W. 1977, Journal of Physical and Chemical Reference Data, 6, 831

\bibitem[{{Feldman}(1992)}]{fel92b}
{Feldman}, U. 1992, \physscr, 46, 202

\bibitem[{{Feldman} {et~al.}(1974){Feldman}, {Doschek}, {Nagel}, {Cowan}, \&
  {Whitlock}}]{fel74}
{Feldman}, U., {Doschek}, G.~A., {Nagel}, D.~J., {Cowan}, R.~D., \& {Whitlock},
  R.~R. 1974, \apj, 192, 213

\bibitem[{{Gabriel}(1972)}]{gab72}
{Gabriel}, A.~H. 1972, \mnras, 160, 99

\bibitem[{{Landi} {et~al.}(2006){Landi}, {Del Zanna}, {Young}, {Dere}, {Mason},
  \& {Landini}}]{lan06}
{Landi}, E., {Del Zanna}, G., {Young}, P.~R., {Dere}, K.~P., {Mason}, H.~E., \&
  {Landini}, M. 2006, \apjs, 162, 261

\bibitem[{{Lemen} {et~al.}(1984){Lemen}, {Phillips}, {Cowan}, {Hata}, \&
  {Grant}}]{lem84}
{Lemen}, J.~R., {Phillips}, K.~J.~H., {Cowan}, R.~D., {Hata}, J., \& {Grant},
  I.~P. 1984, \aap, 135, 313

\bibitem[{{Mann}(1983)}]{man83}
{Mann}, J.~B. 1983, Atomic Data and Nuclear Data Tables, 29, 407

\bibitem[{{Merts} {et~al.}(1976){Merts}, {Cowan}, \& {Magee}}]{mer76}
{Merts}, A.~L., {Cowan}, R.~D., \& {Magee}, Jr., N.~H. 1976, {Los Alamos Report
  UC-20 (LA-6220-MS)}, Tech. rep., Los Alamos Scientific Laboratory, N.M.

\bibitem[{{Phillips} {et~al.}(2005){Phillips}, {Feldman}, \& {Harra}}]{phi05}
{Phillips}, K.~J.~H., {Feldman}, U., \& {Harra}, L.~K. 2005, \apj, 634, 641

\bibitem[{{Phillips} {et~al.}(2008){Phillips}, {Feldman}, \& {Landi}}]{phi08}
{Phillips}, K.~J.~H., {Feldman}, U., \& {Landi}, E. 2008, {Ultraviolet and
  X-ray Spectroscopy of the Solar Atmosphere} (Cambridge University Press)

\bibitem[{{Rice} {et~al.}(2017){Rice}, {Fournier}, {Safronova}, {Goetz},
  {Gutmann}, {Hubbard}, {Irby}, {LaBombard}, {Marmar}, \& {Terry}}]{rice17}
{Rice}, J.~E., {Fournier}, K.~B., {Safronova}, U.~I., {Goetz}, J.~A.,
  {Gutmann}, S., {Hubbard}, A.~E., {Irby}, J., {LaBombard}, B., {Marmar},
  E.~S., \& {Terry}, J.~L. 2017, Unpublished preprint

\bibitem[{{Rice} {et~al.}(2015){Rice}, {Reinke}, {Ashbourn}, {Gao}, {Bitter},
  {Delgado-Aparicio}, {Hill}, {Howard}, {Hughes}, \& {Safronova}}]{rice15}
{Rice}, J.~E., {Reinke}, M.~L., {Ashbourn}, J.~M.~A., {Gao}, C., {Bitter}, M.,
  {Delgado-Aparicio}, L., {Hill}, K., {Howard}, N.~T., {Hughes}, J.~W., \&
  {Safronova}, U.~I. 2015, Journal of Physics B Atomic Molecular Physics, 48,
  144013

\bibitem[{{Rice} {et~al.}(2014){Rice}, {Reinke}, {Ashbourn}, {Gao}, {Victora},
  {Chilenski}, {Delgado-Aparicio}, {Howard}, {Hubbard}, {Hughes}, \&
  {Irby}}]{rice14}
{Rice}, J.~E., {Reinke}, M.~L., {Ashbourn}, J.~M.~A., {Gao}, C., {Victora},
  M.~M., {Chilenski}, M.~A., {Delgado-Aparicio}, L., {Howard}, N.~T.,
  {Hubbard}, A.~E., {Hughes}, J.~W., \& {Irby}, J.~H. 2014, J. Phys. B Atom.
  Mol. Phys., 47, 075701

\bibitem[{{Safronova} \& {Lisina}(1979)}]{saf79}
{Safronova}, U.~I., \& {Lisina}, T.~G. 1979, Atomic Data and Nuclear Data
  Tables, 24, 49

\bibitem[{{Sanchez del Rio} \& {Dejus}(2004)}]{xop04a}
{Sanchez del Rio}, M., \& {Dejus}, R.~J. 2004, in \procspie, Vol. 5536,
  Advances in Computational Methods for X-Ray and Neutron Optics, ed.
  M.~{Sanchez del Rio}, 171--174

\bibitem[{{Seely} \& {Doschek}(1989)}]{see89}
{Seely}, J.~F., \& {Doschek}, G.~A. 1989, \apj, 338, 567

\bibitem[{{Siarkowski} \& {Falewicz}(2004)}]{sia04}
{Siarkowski}, M., \& {Falewicz}, R. 2004, \aap, 428, 219

\bibitem[{{Siarkowski} {et~al.}(2016){Siarkowski}, {Sylwester}, {B{\c
  a}ka{\l}a}, {Szaforz}, {Kowali{\'n}ski}, {Kordylewski}, {P{\l}ocieniak},
  {Podg{\'o}rski}, {Sylwester}, {Trzebi{\'n}ski}, {St{\c e}{\'s}licki},
  {Phillips}, {Dudnik}, {Kurbatov}, {Kuznetsov}, {Kuzin}, \&
  {Zimovets}}]{sia16}
{Siarkowski}, M., {Sylwester}, J., {B{\c a}ka{\l}a}, J., {Szaforz}, {\.Z}.,
  {Kowali{\'n}ski}, M., {Kordylewski}, Z., {P{\l}ocieniak}, S.,
  {Podg{\'o}rski}, P., {Sylwester}, B., {Trzebi{\'n}ski}, W., {St{\c
  e}{\'s}licki}, M., {Phillips}, K.~J.~H., {Dudnik}, O.~V., {Kurbatov}, E.,
  {Kuznetsov}, V.~D., {Kuzin}, S., \& {Zimovets}, I.~V. 2016, Experimental
  Astronomy, 41, 327

\bibitem[{{Sylwester} {et~al.}(2013){Sylwester}, {Phillips}, {Sylwester}, \&
  {K{\c e}pa}}]{bsyl13}
{Sylwester}, B., {Phillips}, K.~J.~H., {Sylwester}, J., \& {K{\c e}pa}, A.
  2013, \solphys, 283, 453

\bibitem[{{Sylwester} {et~al.}(2015{\natexlab{a}}){Sylwester}, {Phillips},
  {Sylwester}, \& {K{\c e}pa}}]{bsyl15}
---. 2015{\natexlab{a}}, \apj, 805, 49

\bibitem[{{Sylwester} {et~al.}(2011){Sylwester}, {Phillips}, {Sylwester}, \&
  {Kuznetsov}}]{bsyl11}
{Sylwester}, B., {Phillips}, K.~J.~H., {Sylwester}, J., \& {Kuznetsov}, V.~D.
  2011, \apj, 738, 49

\bibitem[{{Sylwester} {et~al.}(2005){Sylwester}, {Gaicki}, {Kordylewski},
  {Kowali{\'n}ski}, {Nowak}, {P{\l}ocieniak}, {Siarkowski}, {Sylwester},
  {Trzebi{\'n}ski}, {Baka{\l}a}, {Culhane}, {Whyndham}, {Bentley}, {Guttridge},
  {Phillips}, {Lang}, {Brown}, {Doschek}, {Kuznetsov}, {Oraevsky}, {Stepanov},
  \& {Lisin}}]{jsyl05}
{Sylwester}, J., {Gaicki}, I., {Kordylewski}, Z., {Kowali{\'n}ski}, M.,
  {Nowak}, S., {P{\l}ocieniak}, S., {Siarkowski}, M., {Sylwester}, B.,
  {Trzebi{\'n}ski}, W., {Baka{\l}a}, J., {Culhane}, J.~L., {Whyndham}, M.,
  {Bentley}, R.~D., {Guttridge}, P.~R., {Phillips}, K.~J.~H., {Lang}, J.,
  {Brown}, C.~M., {Doschek}, G.~A., {Kuznetsov}, V.~D., {Oraevsky}, V.~N.,
  {Stepanov}, A.~I., \& {Lisin}, D.~V. 2005, \solphys, 226, 45

\bibitem[{{Sylwester} {et~al.}(2015{\natexlab{b}}){Sylwester}, {Kordylewski},
  {P{\l}ocieniak}, {Siarkowski}, {Kowali{\'n}ski}, {Nowak}, {Trzebi{\'n}ski},
  {{\'S}t{\c e}{\'s}licki}, {Sylwester}, {Sta{\'n}czyk}, {Zawerbny}, {Szaforz},
  {Phillips}, {F{\'a}rn{\'{\i}}k}, \& {Stepanov}}]{jsyl15}
{Sylwester}, J., {Kordylewski}, Z., {P{\l}ocieniak}, S., {Siarkowski}, M.,
  {Kowali{\'n}ski}, M., {Nowak}, S., {Trzebi{\'n}ski}, W., {{\'S}t{\c
  e}{\'s}licki}, M., {Sylwester}, B., {Sta{\'n}czyk}, E., {Zawerbny}, R.,
  {Szaforz}, {\.Z}., {Phillips}, K.~J.~H., {F{\'a}rn{\'{\i}}k}, F., \&
  {Stepanov}, A. 2015{\natexlab{b}}, \solphys, 290, 3683

\bibitem[{{Sylwester} {et~al.}(1998){Sylwester}, {Lemen}, {Bentley}, {Fludra},
  \& {Zolcinski}}]{jsyl98}
{Sylwester}, J., {Lemen}, J.~R., {Bentley}, R.~D., {Fludra}, A., \&
  {Zolcinski}, M.-C. 1998, \apj, 501, 397

\bibitem[{{Sylwester} {et~al.}(2010{\natexlab{a}}){Sylwester}, {Sylwester},
  {Phillips}, \& {Kuznetsov}}]{jsyl10b}
{Sylwester}, J., {Sylwester}, B., {Phillips}, K.~J.~H., \& {Kuznetsov}, V.~D.
  2010{\natexlab{a}}, \apj, 720, 1721

\bibitem[{{Sylwester} {et~al.}(2010{\natexlab{b}}){Sylwester}, {Sylwester},
  {Phillips}, \& {Kuznetsov}}]{jsyl10a}
---. 2010{\natexlab{b}}, \apj, 710, 804

\bibitem[{{Sylwester} {et~al.}(2012){Sylwester}, {Sylwester}, {Phillips}, \&
  {Kuznetsov}}]{jsyl12}
---. 2012, \apj, 751, 103

\bibitem[{{Vainshtein} \& {Safronova}(1978)}]{vai78}
{Vainshtein}, L.~A., \& {Safronova}, U.~I. 1978, Atomic Data and Nuclear Data
  Tables, 21, 49

\bibitem[{{Whiteford} {et~al.}(2001){Whiteford}, {Badnell}, {Ballance},
  {O'Mullane}, {Summers}, \& {Thomas}}]{whiteford01}
{Whiteford}, A.~D., {Badnell}, N.~R., {Ballance}, C.~P., {O'Mullane}, M.~G.,
  {Summers}, H.~P., \& {Thomas}, A.~L. 2001, Journal of Physics B Atomic
  Molecular Physics, 34, 3179

\end{thebibliography}

\end{document}